# Robust Statistical Tests of Dragon-Kings beyond Power Law Distributions

V.F. Pisarenko [1] and D. Sornette [2]

[1] International Institute of Earthquake Prediction Theory and Mathematical Geophysics

Russian Ac. Sci., Profsoyuznaya 84/32, Moscow 117997, Russia

[2] ETH Zurich, Department of Management, Technology and Economics

Kreuzplatz 5, CH-8032 Zurich, Switzerland

Emails: pisarenko@yasenevo.ru and dsornette@ethz.ch

*Abstract*: We ask the question whether it is possible to diagnose the existence of "Dragon-Kings" (DK), namely anomalous observations compared to a power law background distribution of event sizes. We present two new statistical tests, the U-test and the DK-test, aimed at identifying the existence of even a single anomalous event in the tail of the distribution of just a few tens of observations. The DK-test in particular is derived such that the p-value of its statistic is independent of the exponent characterizing the null hypothesis. We demonstrate how to apply these two tests on the distributions of cities and of agglomerations in a number of countries. We find the following evidence for Dragon-Kings: London in the distribution of city sizes of Great Britain; Moscow and St-Petersburg in the distribution of city sizes in the Russian Federation; and Paris in the distribution of agglomeration sizes in France. True negatives are also reported, for instance the absence of Dragon-Kings in the distribution of cities in Germany.

## 1. Introduction

The notion of "outlier" is well known in mathematical statistics and in applied statistics. An outlier is an outstanding observation that occurs in a statistical sample (either in the positive or negative direction for one-dimensional variables or in any direction for multi-dimensional ones). Usually, outliers do not present any interest for the researcher and the statistical problem consists in identifying the outliers (and then throwing them away), whereas the rest of the sample is subjected to the relevant statistical analysis. The outliers are supposed to be mistakes, spurious contaminations and so on, and it is thus best to throw them away so that the remaining sample is free of mistaken observations. Outliers can result from recording or registration errors or can be caused by some other reasons not connected with the physical or social process(es) generating the sample.

Sometimes, the mechanisms generating the outliers (as well as the outliers themselves) are of particular interest and can be subjects of a special discussion and possibly detailed analysis. In such cases, the statistical model generating the time series of observations can be thought of as follows. Some stationary generator (physical or social mechanism) of observations undergoes from time to time (rarely) a change of regime, or an amplification that leads to the production of an unusual rate extreme observation that goes out of the range of the previous observations. The unusual extreme event could also be due to an impact associated with strong external forces or of extreme parameter excursions of the governing mechanism. This observation can be named a **Dragon-King (DK)** [Sornette, 2009]. Such DK can be of interest and even of a very special interest for their special status, significance and for they may reveal previously unknown mechanisms (this is related to the Knightian unknown uncertainty problem [Knight, 1921]). In such cases, the DK should be carefully selected (by a special statistical technique) from the sample for further study. Of course, DK are rare (by definition) and it is not easy to collect a good number of DK for reliable statistical analysis. For instance, aggregating DK from different time series can help.

The purpose of the present paper is to suggest two novel statistical techniques, called respectively the U-test and the DK-test, that specifically aim at identifying or diagnosing the existence of even a single Dragon-King (DK) in a finite sample. Section 2 presents the two tests. Section 3 (respectively section 4) applies them to the distributions of cities (respectively agglomerations) in a number of countries. Section 5 concludes.

## 2. Statistical technique for Dragon-King tail analysis

We present two statistical tools, respectively referred to as the DK-test and the U-test.

### *2.1 The DK-test.*

Let us consider a sample of independently indentically distributed (iid) random variables (rv) $x_1, ..., x_n$ with exponential probability density function (PDF)

$$f(x) = a \cdot exp(-ax), \qquad a > 0, \quad x \geq 0. \qquad (1)$$

We could assume as well the Pareto PDF

$$f(x) = b \cdot h^b / x^{1+b}, \qquad b > 0, h > 0, \quad x \geq h. \qquad (2)$$

All our further considerations are valid for the Pareto distribution, since the natural logarithm of normalized Pareto-rv $X/h$ has the exponential distribution (1) with $a = b$.

We order the sample $x_1, ..., x_n$ and get $x_{1,n} \geq x_{2,n} \geq ... \geq x_{n,n}$ (the convention is that rank 1 is the largest, rank 2 is the second largest, and so on). We suspect that some number $r$ of the highest ranks

$$x_{r,n} \leq x_{r-1,n} \leq ... \leq x_{1,n} \qquad (3)$$

could be generated by some distinct heavy tail distribution (the DK regime), whereas all the rest of the rv

$$x_{n,n} \leq x_{n-1,n} \leq \ldots \leq x_{r+1,n} \tag{4}$$

are generated by the exponential distribution (1).

The number $r$ is fixed a-priori from some preliminary study ($r \leq n - 1$). In practice, as we shall see, we can start with r=1 to test whether the largest rv is a generated by a process different from the rest, then go to r=2 to test if the two largest events are generated by a distribution distinct from the exponential distribution (1), and so on. Note that, in the case where the bulk of the sample is generated by the Pareto distribution (2), our DK-test corresponds to testing if the largest r events are produced by a process with an even heavier tail, for instance a singular measure concentration at extremes values, in the spirit of the extension of decision theory involving the replacement of the monotone continuity axiom into purely finitely additive measures that focus on extreme events [Chichilnisky, 2000; 2009].

Our goal is to construct a test for the null hypothesis:

***$H_0$: all observations of the sample are generated by the same exponential distribution (1).***

An alternative to $H_0$ consists in the existence of Dragon-Kings, defined as very large observations generated by some different distribution with a heavy tail.

Let us consider the *spacings* $y_k$ defined as

$$y_k = x_{k,n} - x_{k-1,n}, \quad k = 1,\ldots n\text{-}1 ; \tag{5a}$$

$$y_n = x_{n,n} . \tag{5b}$$

It can be proved (see Embrechts et al. Chapter 4, Section 4.1 Order Statistics, example 4.1.5) that *if $x_1,\ldots, x_n$ are iid rv with PDF (1), then the spacings $y_1, y_2,\ldots, y_n$ are independent, exponentially distributed and $y_k$ has mean value $1/(a \cdot k)$).*

We construct the following test statistic $T$. Defining the rv

$$z_k = k \cdot y_k, \quad k=1,\ldots,n , \tag{6}$$

the proposed test statistic reads

$$T = \frac{\frac{1}{r}(z_1 + \ldots + z_r)}{\frac{1}{(n-r)}(z_{r+1} + \ldots + z_n)} . \tag{7}$$

Since the rv $y_1, y_2,\ldots, y_n$ are exponentially distributed, the rv $z_k$ for $k=1,\ldots,n$ are also exponentially distributed. But an exponential distribution is, apart from a constant factor, nothing but the $\chi^2$-distribution with 2 degrees of freedom. The numerator of (7) is the sum of $r$ rv variables, each of them being distributed with the $\chi^2$-distribution with 2 degrees of freedom. Hence, the numerator is distributed (apart from a constant factor) according to the $\chi^2$-distribution with *2r* degrees of freedom. Similarly, the sum in the denominator of (7) has (apart from the same constant factor as in the numerator) a $\chi^2$-distribution with *2(n-r)* degrees of freedom. It follows that the statistic *T* is distributed according to the *f*-distribution with *(2r, 2(n-r))* degrees of freedom. The corresponding *p*-value for the hypothesis $H_0$ is thus given by

$$p = 1 - F(T, 2r, 2(n-r)), \qquad (8)$$

where *F(.,2r, 2(n-r))* denotes the cumulative distribution function (CDF) of the *f*-distribution with *(2r, 2(n-r))* degrees of freedom. It should be stressed that the *p*-value given by (8) does not depend on the exponential parameter *a* (or, the Pareto parameter *b*).

In constructing the test statistic T, there is one potentially non-robust operation, which is the construction of the rv $z_k$ according to expression (6). If we take large *n* values (e.g. all available observations), then small deviations in $y_k$ can result in large deviations of $z_k$ for the largest ranks, that is, the smallest deviations $y_k$. This non-robust operation thus mainly impacts the small ranks corresponding to small rv and is thus not a serious problem as the *T*-test focuses on the tail behavior. However, it seems safer to restrict *n* to, say, *20-30*.

In addition to the analysis of Dragon-Kings, it should be remarked that the T-test can also be applied to the detection of such phenomena as "clipping" maxima, or "bent down" of distributions. Such events, that we refer to as "negative Dragon-Kings" would correspond to smaller spacing in the tail range, that deviate negatively (being smaller events than expected) from the extrapolation of the exponential distribution (1) (or power law (2)) taken as the null hypothesis for the generating mechanism.

In the presence of Dragon-Kings, the numerator in the statistic (7) is on average larger than the denominator under alternative hypotheses, making *T* typically larger than *1*. In contrast, for "bent downs" (deviations from the null from below), the numerator is on average smaller than the denominator for alternative hypotheses, making *T* typically smaller than *1*.

A last remark is in order. In the presence of a single DK in the sample, the other n-1 normalized spacing $z_k$ exhibit a regular behavior described by the null PDF. When two or more DK exist, the T-test performance will depend on the relationship between the different DK and their level of inter-dependence. In other words, the diagnostic of DK needs to take into account the length of the upper tail that serves as a background. This effect will be investigated practically as we present below the first tests on empirical data.

## *2.2 The U-test.*

The U-test consists in a modification of the test introduced by Pisarenko and Sornette [2004], in which the parameter estimation is performed in an arbitrary interval of the random variable in question, i.e. with truncations both from above and from below.

We use the exponential cumulative distribution function (CDF) *F(x| b)*, depending on the unknown parameter *b* : *F(x| b) = 1 – exp(-b·x)*. As an illustration and for the sake of concreteness, we apply this CDF to the logarithms of the initial rv consisting in city sizes, as well as to agglomeration sizes of a given country. In other words, we fit the exponential CDF to the

log-sizes. Note that this corresponds to a null hypothesis according to which the distribution of sizes is a power law with the same exponent *b* (see Malevergne et al. (2011) for a recent review of the literature and new light on the controversies concerning the distribution of city sizes in the US).

We use the Maximum Likelihood (ML) estimates derived for samples truncated by a lower threshold *h* and restricted from above by one or several highest ranks of the ordered sample, chosen preliminarily. The choice of the threshold *h* and of the highest restricting rank is not fully formalized but hopefully the final result depends weakly on this choice. The algorithm to process the initial sample $w_1,...,w_N$ works as follows.

1. By visual inspection, we determine the lower threshold *h* and pick out only observations exceeding *h*. We denote the ordered sample of this selected set as $v_{1,n} \geq ... \geq v_{n,n}$.
2. We take natural logarithms, leading to the new rv $x_{k,n} = log(v_{k,n})$.
3. By visual inspection, we determine *r* largest observations that can possibly contain one or more DK, where *r* can be zero or some positive integer number. This corresponds to assuming that our sample can contain some number of DK, which is not larger than *r*.
4. We estimate by the Maximum Likelihood method the exponent *b* from the subsample $x_{r+1,n},..., x_{n,n}$. For this, we use the form of the PDF of $x_{r+1,n},..., x_{n,n}$ given by

$$f(x_{r+1,n},..., x_{n,n}) \sim [1- F(x_{r+1,n}| b)]^r \prod_{k=r+1}^{n} f(x_{k,n}|b); \quad F(x|b)=1-exp(-b(x-h)); \quad x \geq h. \quad (9)$$

The ML-estimate of the parameter *b* is denoted as $\hat{b}$.

5. We calculate the p-values of $x_{k,n}$ using the following equation (see below for the derivation):

$$p(x_{k,n}) = 1 – betainc(F(x_{k,n}|\hat{b}), n-k+1, k), \quad k = 1,...,n. \quad (10)$$

where *betainc(. , n-k+1, k)* is the normalized incomplete beta-function. The p-value (10) is defined as the probability of exceeding the observed value $x_{k,n}$ under the null hypothesis of the exponential distribution for the log-sizes. From the *p*-values, we can judge about the observations $x_{r-1},..., x_{1,n}$ suspected as being DKs. If among the *p*-values $p(x_{r-1,n}),...,p(x_{1,n})$, there exist some small values (say, less than 0.10), then one can conclude that there are some DK among the log-sizes $x_{r-1,n},..., x_{1,n}$ with individual confidence level *1-p*.

Expression (10) can be derived as follows. If a continuous rv, say *ξ*, is inserted into its own distribution function, then the rv *v=F(ξ)* is distributed uniformly in the interval *(0, 1)*. We apply this rule to the CDF $F(X | b) = 1 – exp(-b·(X-h))$, $(X = log(W))$. If we would know the true parameter value *b* and inserted the rv *X* into the expression of *F (X | b)*, then the obtained random variable would be uniformly distributed in the interval *(0,1)*. Then the *k*-th term of the ordered sample $x_{k,n}$ would have the PDF

$$f_k(u) = \frac{n!}{(n-k)!(k-1)!} u^{k-1}(1-u)^{n-k}; \quad 0 < u < 1 \quad (11)$$

which gives equation (10). But in practice, the parameter $b$ is unknown and should be replaced by its statistical estimate. In this paper, we use the Maximum Likelihood estimate $\hat{b}$.

Inserting this estimate $\hat{b}$ into the expression $F(X | b) = 1 - exp(-b \cdot (X-h))$ gives a random value $F(\xi, \hat{b})$ which is distributed uniformly in the interval $(0, 1)$ only approximately, and the approximation being all the better, the larger is the sample size $n$ and the better is the accuracy of the estimate $\hat{b}$. For each $x_{k,n}$, the probability (or $p$-value) of exceeding $x_{k,n}$ (under the condition that the distribution of the rv $F(X | \hat{b})$ is sufficiently well approximated by the uniform distribution) is given by

$$p(x_{k,n}) = \frac{n!}{(n-k)!(k-1)!} \int_{Z_k}^{1} w^{k-1}(1-w)^{n-k} dw = 1 - betainc(Z_k, n-k+1, k), \quad k=1,...,n, \quad (12)$$

where $betainc(., k, n-k+1)$ is the normalized incomplete beta-function; $Z_k = F(x_{k,n} | \hat{b})$.

Of course, the rv $x_{k,n}$ are not independent. We estimate all corresponding p-values, so to say, individually for each order statistic $x_{k,n}$. The series $p(x_{1,n}),..., p(x_{n,n})$ then characterizes the deviations of the ordered sample $x_{1,n} \geq ... \geq x_{n,n}$ from its "average" positions $n/(n+1), (n-1)/(n+1),..., 1/(n+1)$ in terms of the probabilities of the individual deviations regardless of sample size.

Table 1 provides the interpretation of the $p$-value for the two DK-test and U-test.

| Table 1. Interpretation of different $p$-values for the DK-test and U-test. PL means "power law" | | | | |
|---|---|---|---|---|
| | | $p < 0.10$ | $0.10 < p < 0.90$ | $p > 0.90$ |
| | DK-test testing the $r$ first spacing versus the (n-r) preceding spacing values; | $\leq r$ Dragon-King(s) | no significant deviation from PL distribution | $\leq r$ "negative Dragon-King(s)" |
| | U-test | Dragon-King with respect to fitted Power law distribution | no significant deviation from fitted PL-distribution | "negative Dragon-King" with respect to fitted PL-distribution |

## 3. Testing for Dragon-Kings in city sizes

We apply the DK-test and U-test to the distribution of city sizes of a number of typical countries. The data is taken from the website www.citypopulation.de, which has compiled a large amount of data from various sources. Each country has its own definition what the term city means. Mostly it refers to the smallest administrative unit of a country having a predominately urban population.

### *3.1 Great Britain: evidence of one DK (London)*

Fig.1 shows the complementary cumulative distribution function (CDF) *1-F(x)* in double log-scale for the city sizes in Great Britain in the year 2008. Using the lower threshold *h = 50300* (i.e., cities larger than or equal to 50300 inhabitants), the sample size is *n = 194*. One can observe that a straight line qualifying a power law distribution is a good approximation of the distribution for all cities, except for the largest one (the city of London), which can be suspected as a DK. We thus considered the 193 observations (without London) in order to obtain the ML-estimate of parameter *b* and found *b =1.502*. Note this value is significantly larger than 1, i.e., Zipf's law with *b=1* for large cities (more than 50300 inhabitants) is rejected [Malevergne et al., 2011].

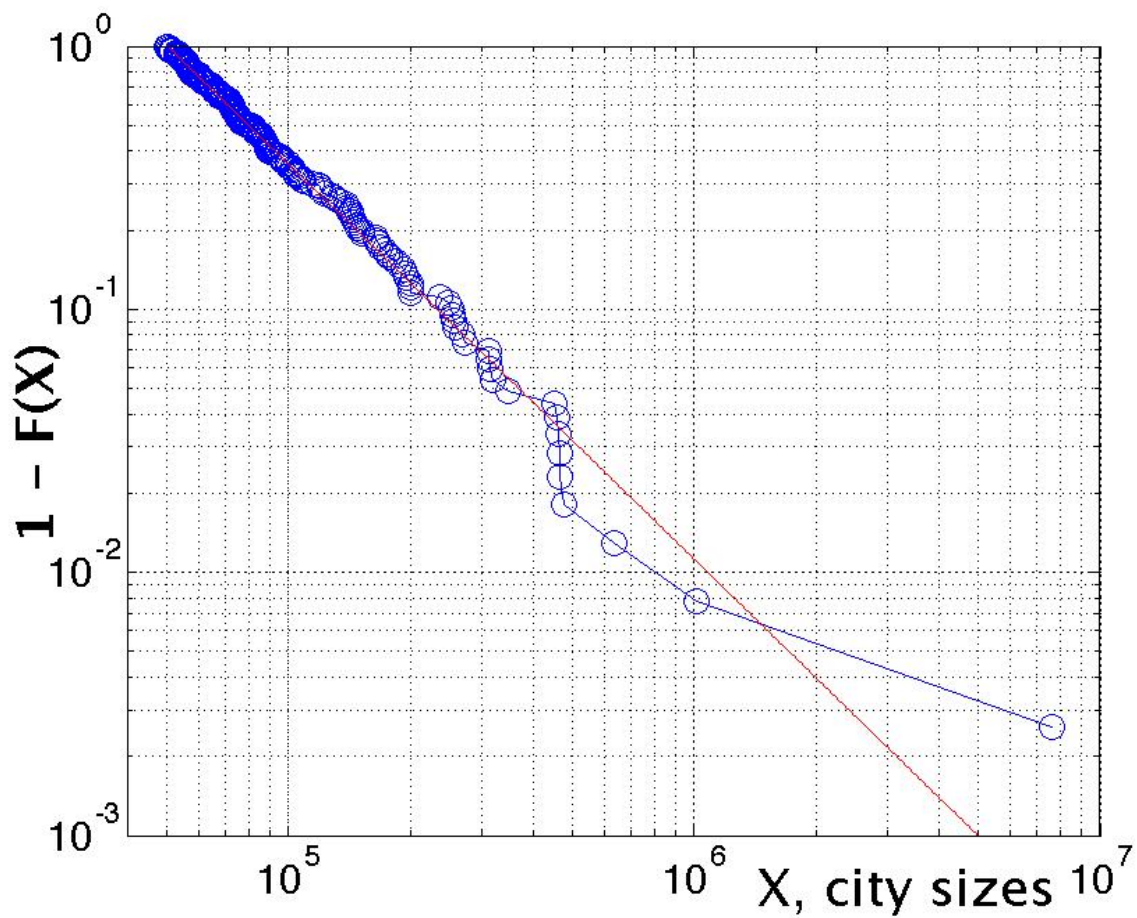

**Figure 1**: Complementary cumulative distribution function of the 194 largest cities in Great Britain with more than 50300 inhabitants. The straight line is the Pareto fit on n=193 (excluding the largest r=1 city), yielding the exponent b=1.502.

To apply the U-test, we inserted $b = 1.502$ into $F(x) = 1 - exp(-b \cdot (x-h))$ and calculated the p-values by equation (10). The result is shown on Fig.2. We see that the largest observation (London) corresponds to $p = 0.098 < 0.10$, given a first marginal indication that that London is a DK.

To apply the DK test, we take the largest spacing $(x_{1,194} - x_{2,194})$ and compared it with the average value of the following ones. Equation (8) with $r=1$ (one DK candidate) and a variable number $n_1 = 2, 3, ...36$ of the first following ranks allows us to construct Fig.3. Comparing the largest spacing $(x_{1,194} - x_{2,194})$ with the average spacing of the $n_1 = 2, 3$ or $4$ preceding spacing leads to p-values larger than 0.10. But from $n_1 = 5$ and higher, all p-values are small than 0.10. Thus, one can conclude that the largest spacing $(x_{1,194} - x_{2,194})$ is formed by a DK, i.e. London is qualified as a DK.

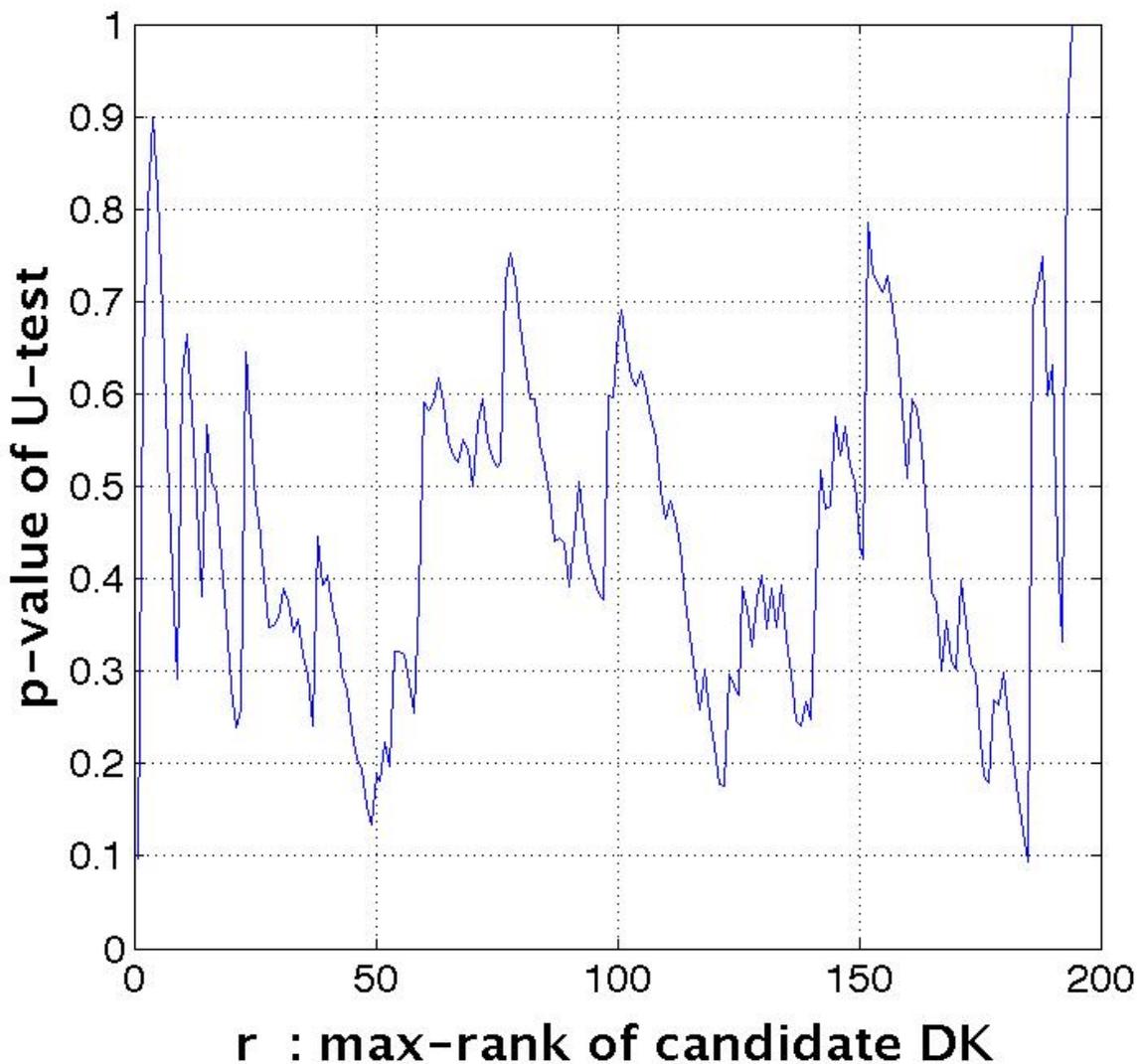

**Figure 2**: *p*-value of the U-test as a function of the maximum rank *r* for the candidate DK (London among the Great Britain cities, 2008).

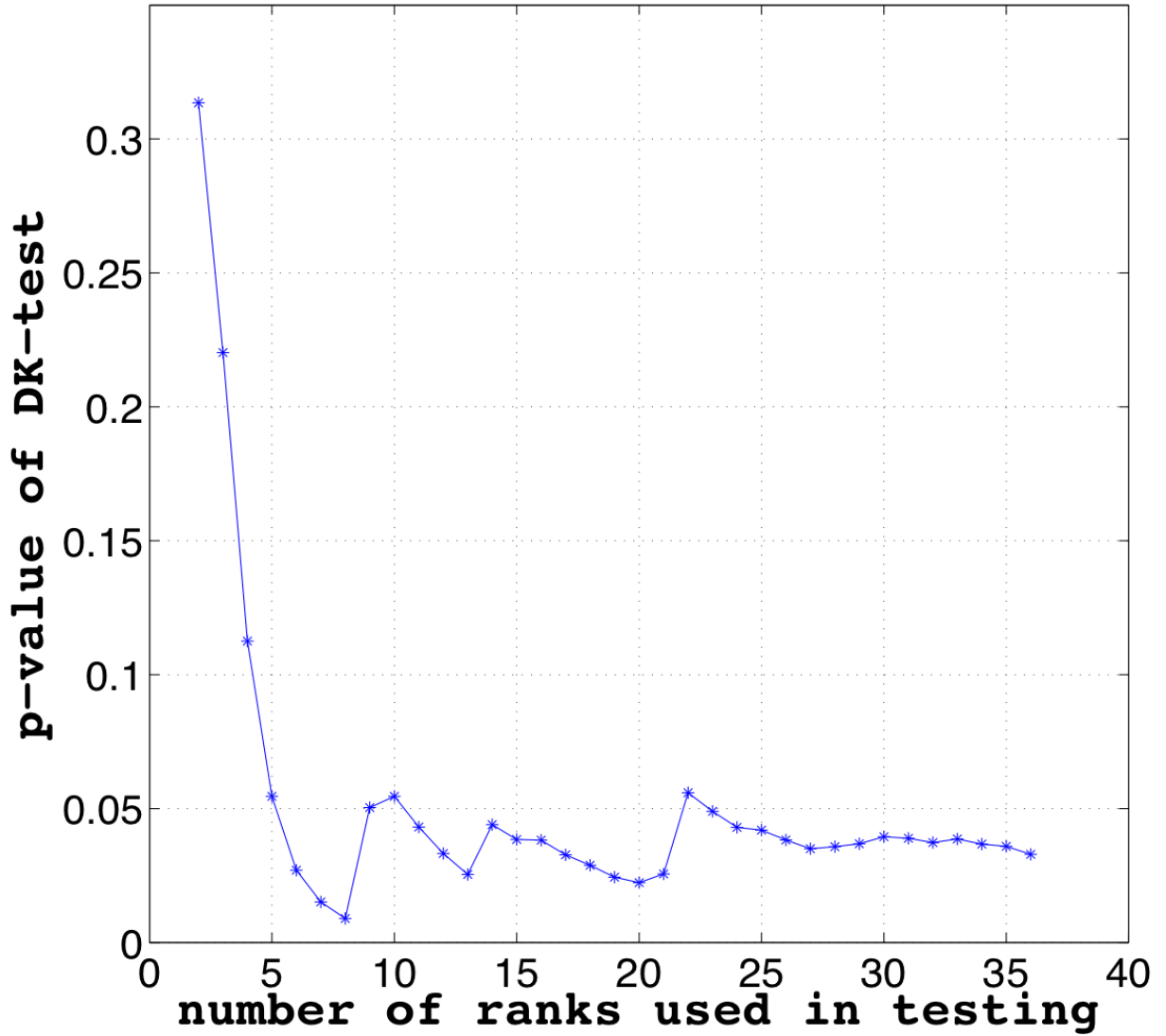

**Figure 3**: *p*-value of the DK-test as a function of the number $n_1$ of ranks used in the testing, for r=1 (Great Britain cities, 2008).

### *3.2 Russian Federation: evidence of two DK (Moscow and Saint-Petersburg)*

Fig.4 shows the complementary cumulative distribution function CCDF *1-F(x)* in double log-scale for Russian city sizes, in 2010, with a total number *n = 116* cities. One can observe that two maximum observations (Moscow and Saint-Petersburg) deviate rather significantly from the others and can be suspected as being DK.

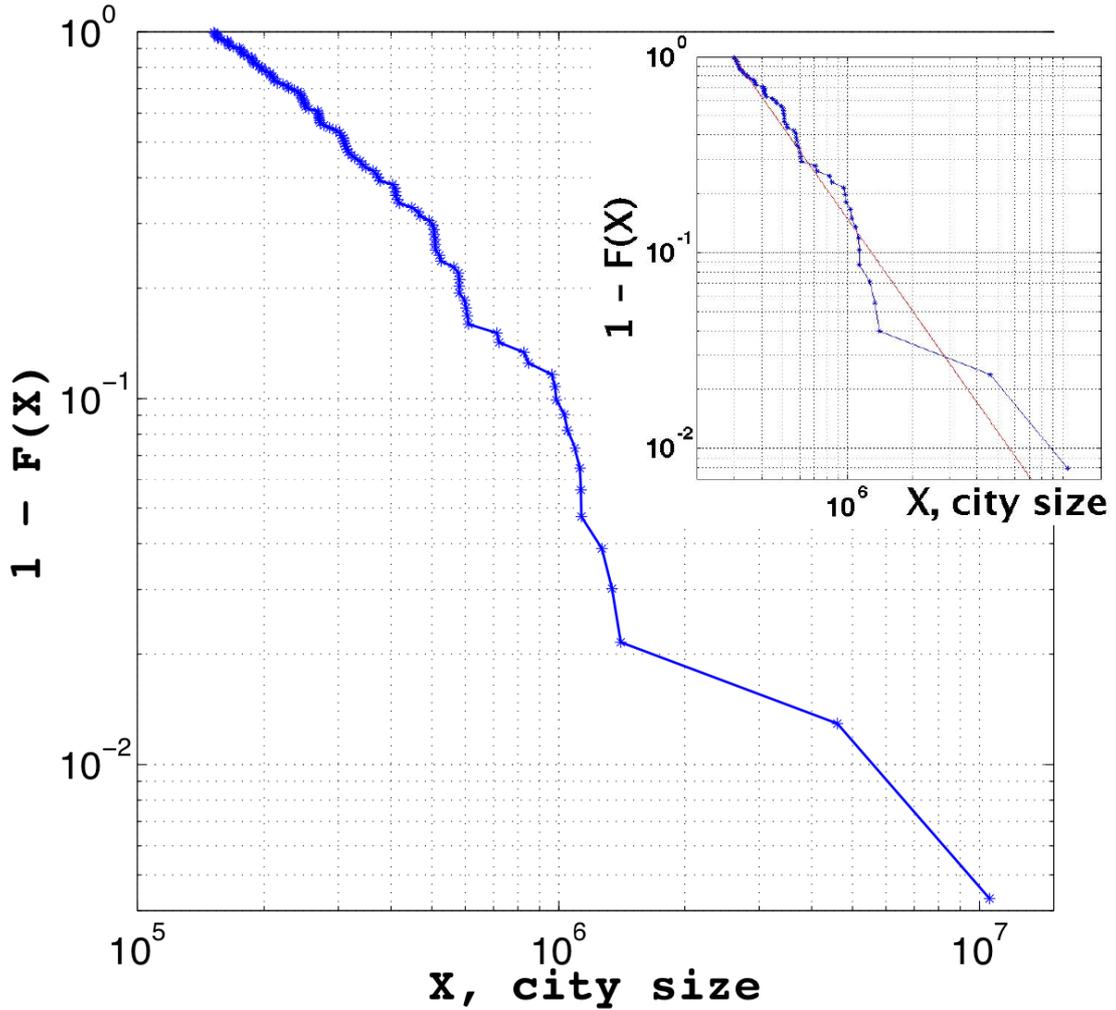

**Figure 4**: Complementary cumulative distribution function (CCDF) of the sizes of the *n=116* largest cities in the Russia Federation (2010). Inset: tail of the CCDF for the *n=61* largest cities with more than *h=296000* inhabitants and fit with a power law (straight line on the graph) whose exponent is found equal to *b=1.56*.

The choice of the lower threshold $h$, beyond which the power law null hypothesis is supposed to hold, is not obvious, because the tail of the CCDF in Fig.4 has an overall negative curvature (except for the two largest data points). We chose the value $h = 296,360$ inhabitants (corresponding to selecting the 63 largest city sizes), as being the size beyond which the asymptotic power law tail holds. Our goal is to fit a power law (straight line fitting in double-logarithmic scale) for all city sizes larger than this threshold, excluding the two largest cities that we consider as candidate DK. Recall that the power law is the null hypothesis against which the DK hypothesis is tested. It should be noted that a straight line approximation of the tail cannot be quite satisfactory in this case, since the negative curvature can be observed to continue till the tail end (excepting for the two largest observations). Perhaps, a log-normal approximation of the tail would be more appropriate here, but then the DK-test defined above would not applicable.

Fitting a power law to the CCDF over the 61 largest observations above the lower threshold $h = 296,360$ (without Moscow and Saint-Petersburg), we obtain a ML-estimate of the parameter $b$ equal to *1.56*. The power law fit corresponding to the straight line is shown in the inset of Figure 4.

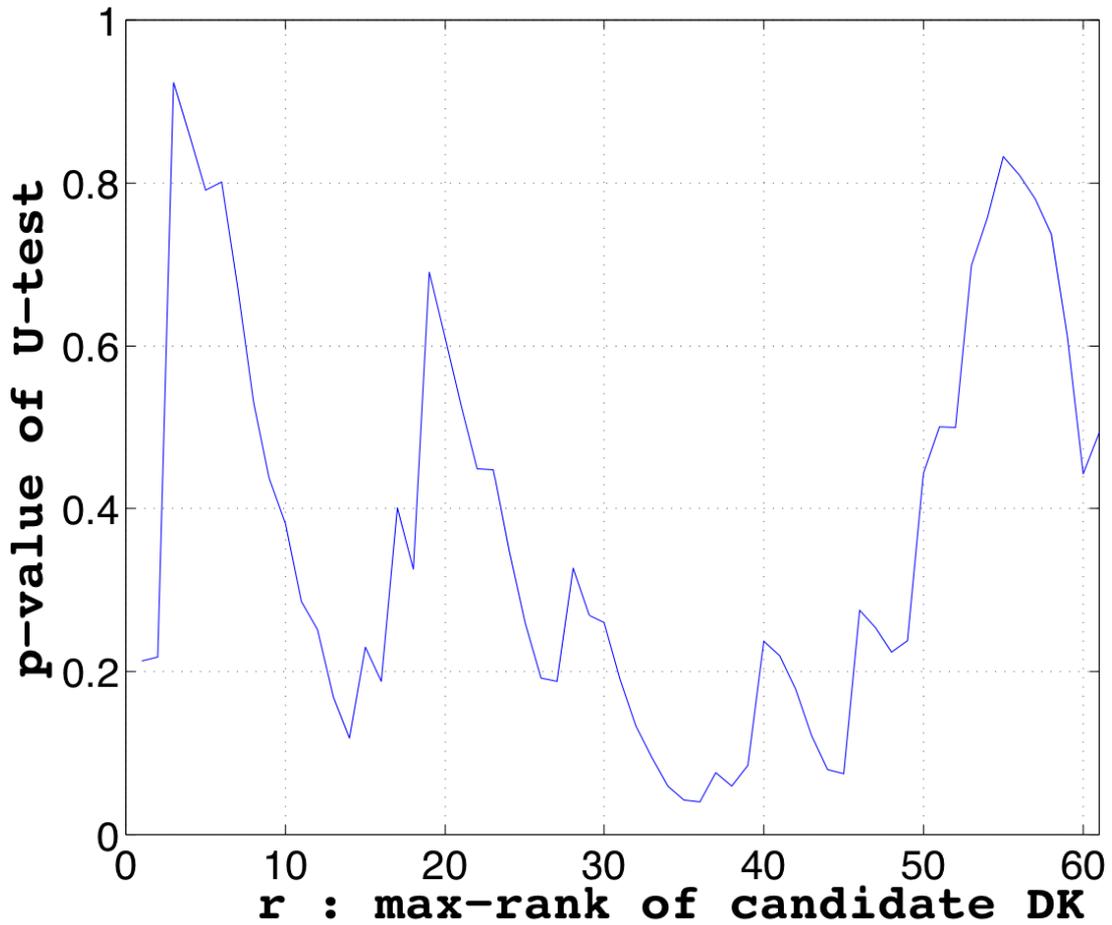

**Figure 5**: *p*-value of the U-test as a function of the maximum rank *r* for candidates DK among the 63 largest observations above the lower threshold *h = 296,360* for cities in the Russia Federation (2010).

To apply the U-test, we inserted *b = 1.56* into *F(x) = 1 − exp(-b·(x-h))* and calculated the p-values by equation (10). The result shown in Fig.5 shows that there are no p-values less than 0.10 till very large ranks *(r = 32-39),* where DK are out of the question. Therefore, one can conclude that U-test did not discover any DK, although some p-values near *r = 14* are close to 0.10*. This rank *r = 14* corresponds to the pronounced downward bend that can be seen in figure 4, suggesting indeed a change of regime.

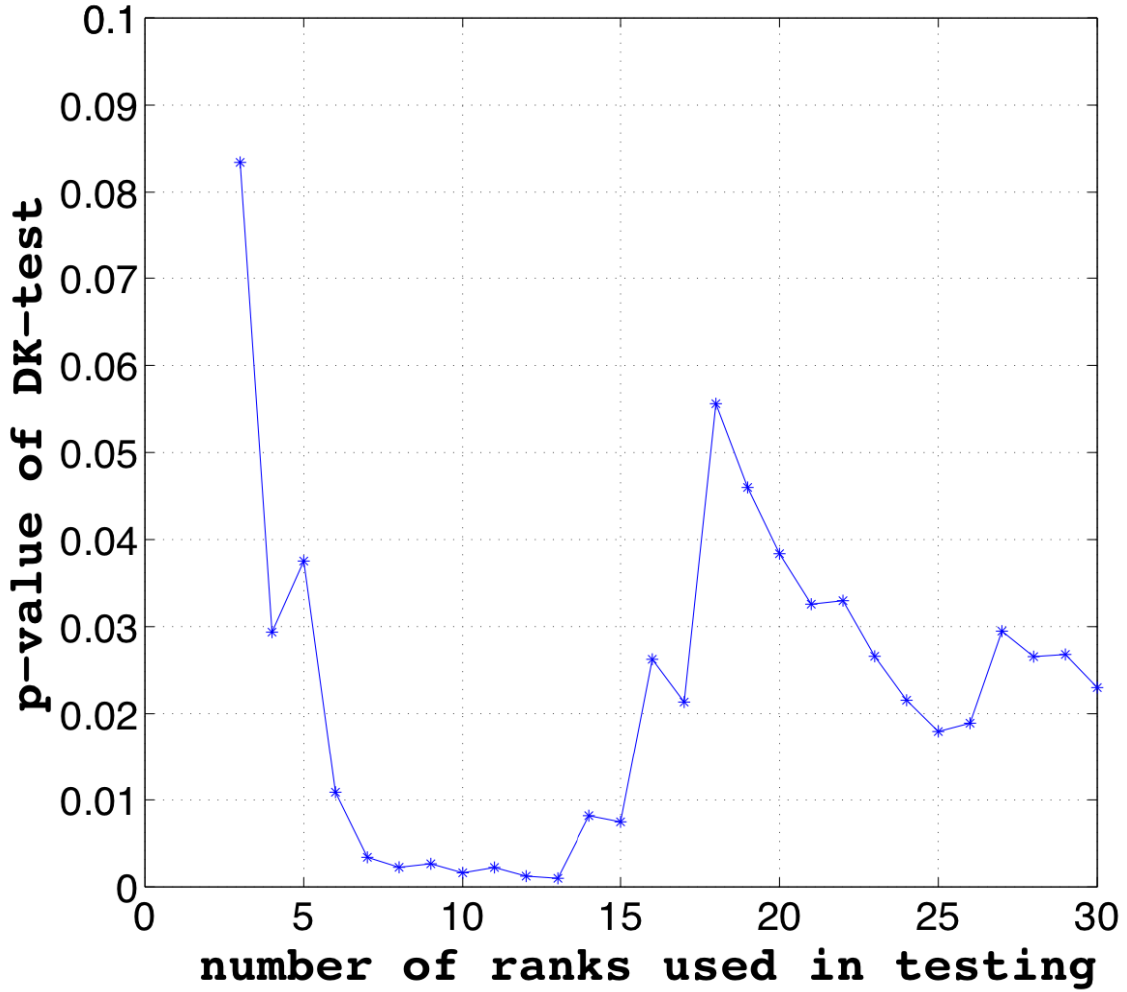

**Figure 6**: *p*-value of the DK-test as a function of the number $n_1$ of ranks used in the testing, for r=2 (Moscow and Saint-Petersburg as DK candidates for the distribution of Russian cities, 2010).

To apply the DK test, we took the two largest spacing $(x_{1,63} - x_{2,63})$, $(x_{2,63} - x_{3,63})$ and compared them with the average value of the following ones. Equation (8) with *r=2* (two DK candidates) and a variable number $n_1 = 2, 3, ...30$ of the first following ranks yields the *p*-value as a function of $n_1$ shown in Fig.6. As the *p*-value is significantly smaller than 0.1, one can conclude that Moscow and Saint-Petersburg can be considered as DK (in accordance with the DK-test).

### *3.3 USA: no evidence of DK*

Fig.7 shows the complementary cumulative distribution function CCDF *1-F(x)* in double log-scale for USA city sizes, in 2009, with a total number *n = 283* cities. There is no obvious deviation from the power law fit shown as the straight line, even for the two largest cities, New York and Los Angeles. For the lower threshold *h*, we take the smallest size *h = 100160* inhabitants among the *283* cities since the power law seems to be adequate over the whole range. The ML-estimate of the parameter *b* is found equal to *1.406*.

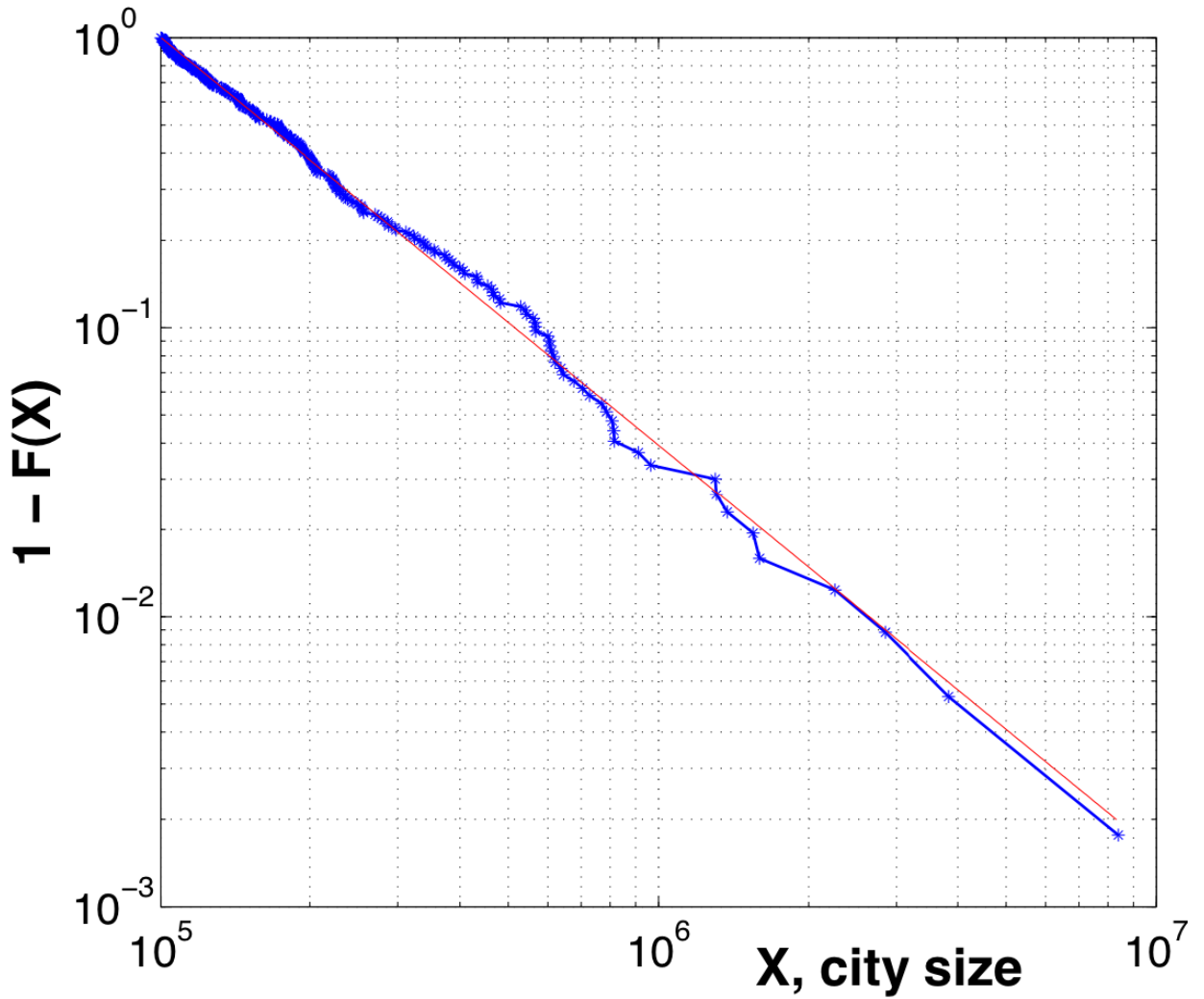

**Figure 7**: Complementary cumulative distribution function (CCDF) of the sizes of the *n=283* largest cities in the United States of America (2009). The straight line shows the power law fit over the whole range with an exponent *b=1.406*.

To apply the U-test, we inserted $b = 1.406$ into $F(x) = 1 - exp(-b \cdot (x-h))$ and calculated the p-values by equation (10). The result in Fig.8 shows that there are no p-values less than 0.10. Therefore, one can conclude that U-test did not discover any DK.

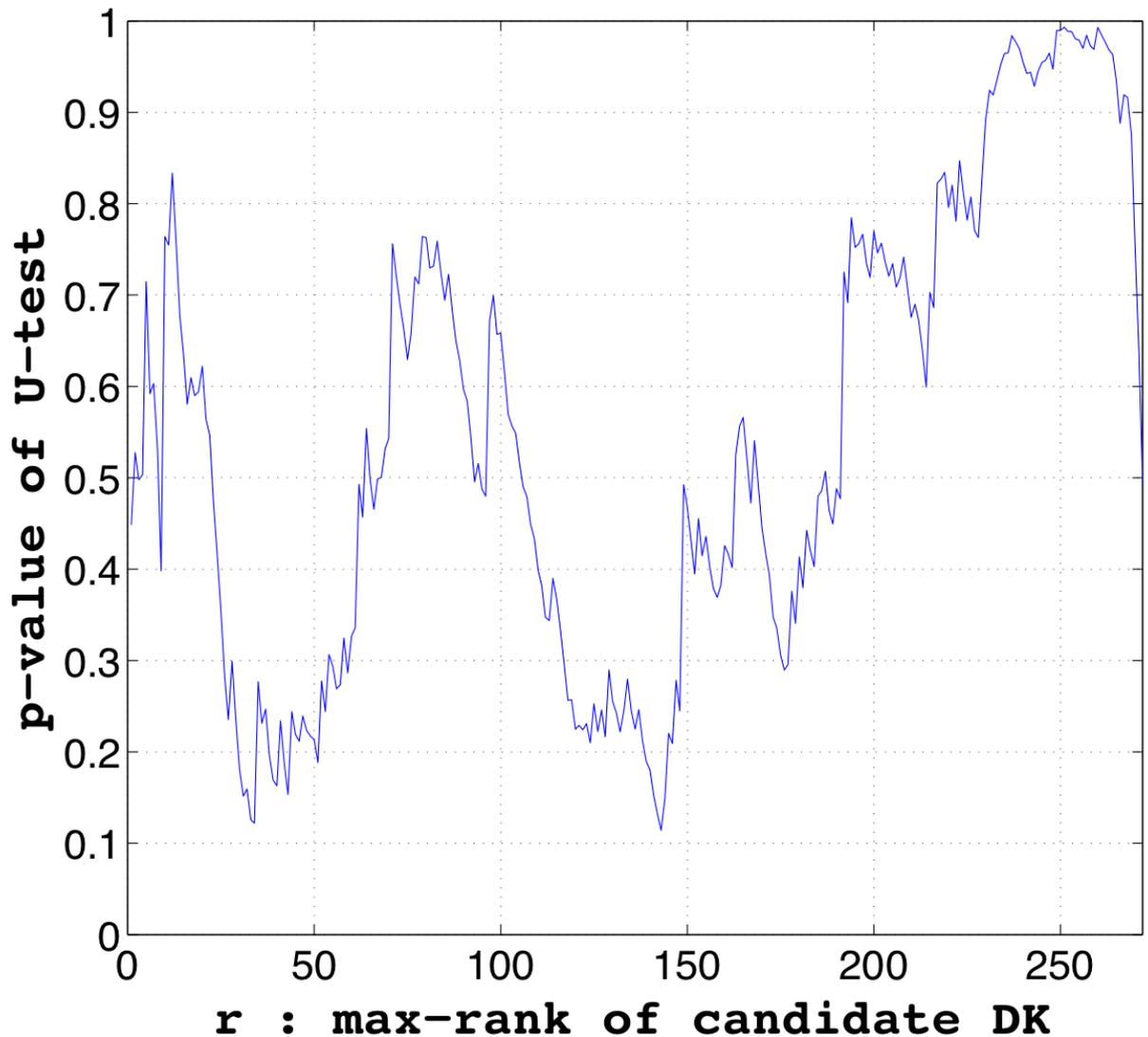

**Figure 8**: *p*-value of the U-test as a function of the maximum rank *r* for candidates DK among the 283 largest observations above the lower threshold *h = 100160* for cities in the USA (2009).

To apply the DK test, we took the largest spacing $(x_{1,283} - x_{2,283})$ and compared it with the average value of the following ones. Equation (8) with *r=1* (one DK candidate) and a variable number $n_1 = 2, 3, ...200$ of the first following ranks yields the *p*-value as a function of $n_1$ shown in Fig.9. As the *p*-values are all larger than 0.1, one can conclude that no observation can be considered as a DK (in accordance with the DK-test). Combined with the U-test, the distribution of the 283 largest USA cities is well described by a pure power law distribution. This confirms for this data set the results obtained by Malevergne et al. (2011) using the UMPU (uniformly most powerful unbiased) test comparing the power law model with the log-normal model.

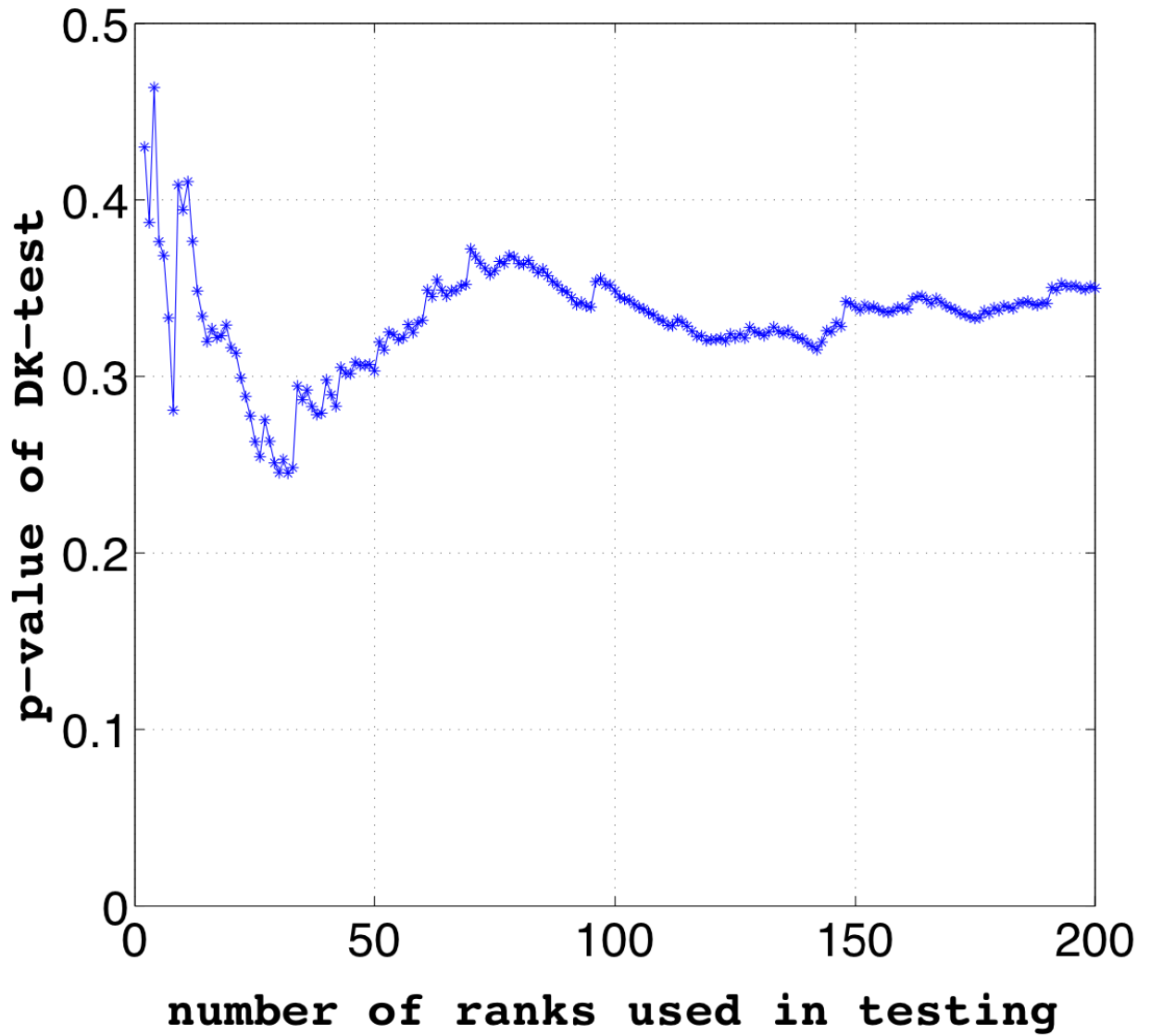

**Figure 9**: *p*-value of the DK-test as a function of the number $n_1$ of ranks used in the testing, for r=1 (New York as DK candidate for the distribution of USA cities, 2009).

*3.4 Germany: no evidence of DK*

Fig.10 shows the complementary cumulative distribution function CCDF *1-F(x)* in double log-scale for German city sizes, in 2008, with a total number *n = 187* cities. There is perhaps one maximum observation (Berlin) that deviates from the others but not significantly. The choice of the lower threshold *h*, beyond which the power law null hypothesis is supposed to hold, is not obvious, because the tail of the CCDF in Fig.10 has an overall negative curvature, a bit similar to the situation for the CCDF of Russian cities shown in Fig.4, and some undulations. We chose the value *h=141000* inhabitants (corresponding to selecting the 55 largest city sizes), as being the size beyond which the asymptotic power law tail holds. Our goal is to fit a power law (straight line fitting in double-logarithmic scale) for all city sizes larger than this threshold. The corresponding power law fit is shown in the inset of Fig.10. The ML-estimate of the power law exponent *b* is *1.321*.

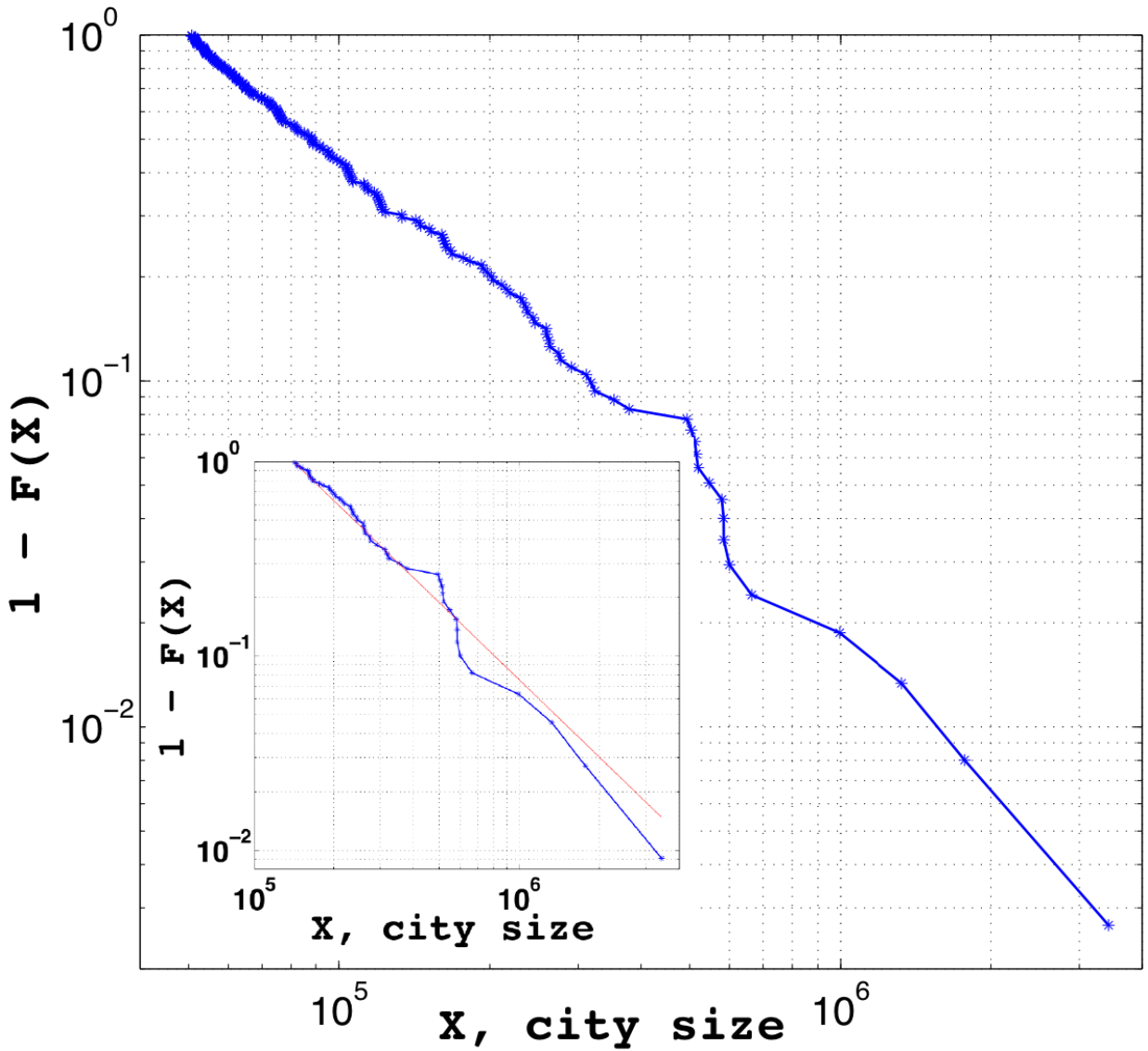

**Figure 10**: Complementary cumulative distribution function (CCDF) of the sizes of the *n=187* largest cities in Germany (2008). Inset: tail of the CCDF for the *n=55* largest cities with more than *h=141000* inhabitants and fit with a power law (straight line on the graph) whose exponent is found equal to *b=1.321*.

To apply the U-test, we inserted the value *b = 1.321* into *F(x) = 1 – exp(-b·(x-h))* and calculated the p-values by equation (10). The result in Fig.11 shows that there is one p-value smaller than 0.10, namely for *r = 15*. This means that the U-test did not find any DK in the largest 14 observations, but still there is some doubt concerning a possible change of regime around rank 15. Note that this rank corresponds to the peak of the largest oscillatory deviations from the power law, as can be seen in Fig.10.

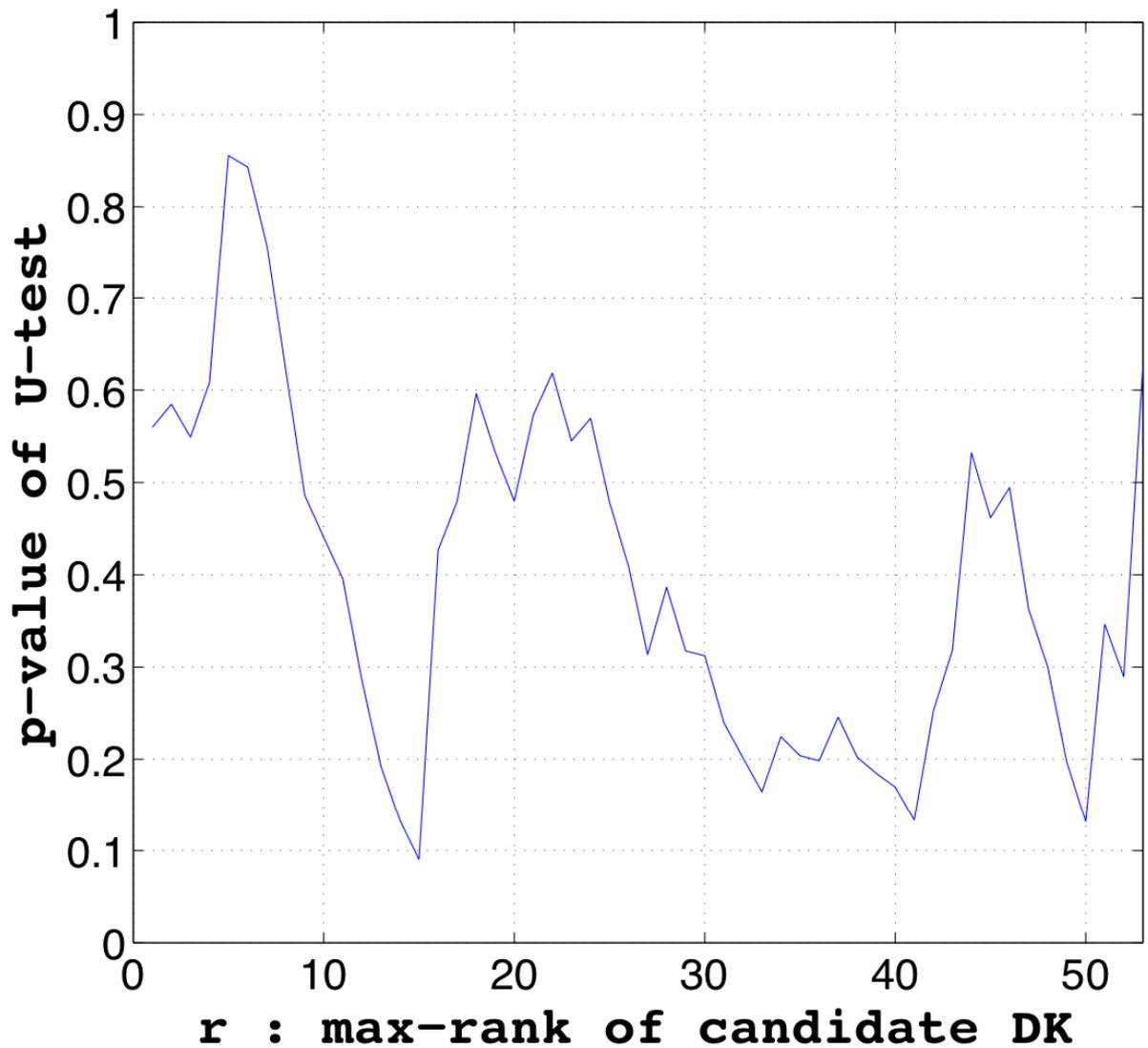

**Figure 11**: *p*-value of the U-test as a function of the maximum rank *r* for candidates DK among the 55 largest observations above the lower threshold *h=141000* for cities in Germany (2008).

To apply the DK test, we took the largest spacing $(x_{1,55} - x_{2,55})$ and compared it with the average value of the following ones. Equation (8) with *r=1* (one DK candidate) and a variable number $n_1$ = *2, 3, ...,55* of the first following ranks yields the *p*-value as a function of $n_1$ shown in Fig.12. As the *p*-values are all larger than 0.1, one can conclude that no observation can be considered as a DK (in accordance with the DK-test).

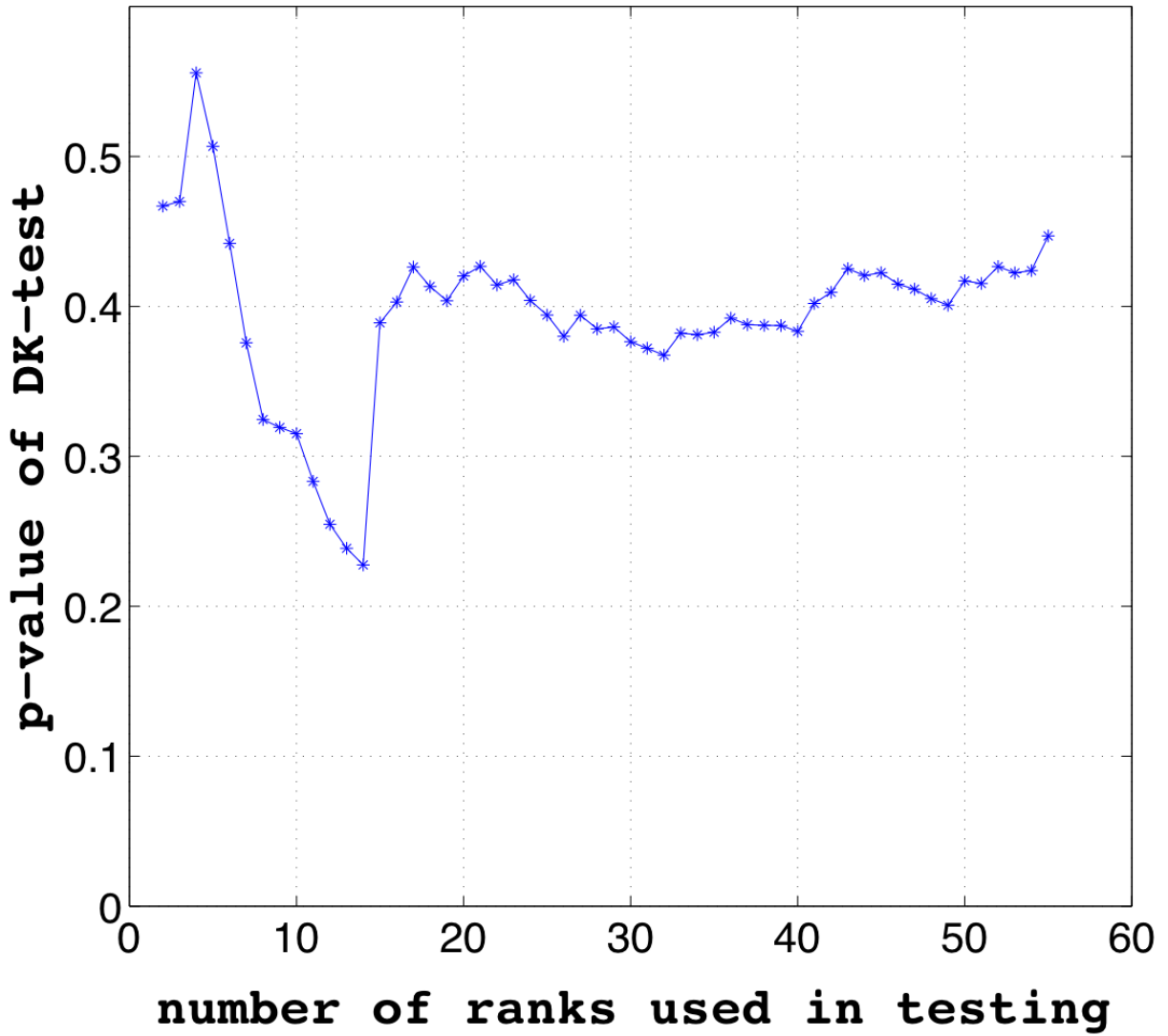

**Figure 12**: *p*-value of the DK-test as a function of the number $n_1$ of ranks used in the testing, for r=1 (Berlin as DK candidate for the distribution of German cities, 2008).

### 3.5 Summary of results for 13 countries

Table 2 provide the main summary statistics and synopsis of the results obtained with the U-test and DK-test applied to the distribution of cities in 13 different countries.

Table 2. City size statistics. m = number of spacing in denominator of (7) giving a p-value smaller than 0.10;  r = number of spacing in the numerator of (7); n = number of cities used in the analysis/

| country | max size | b, Pareto exponent (ranks used for estimation) | min p-value of U-test among first 10 ranks | DK-test; (r; m) giving p < 0.10 |
|---|---|---|---|---|
| Great Britain, n=194 | 7,619,800 | 1.502 (2 ÷ 194) | p= 0.098 (DK London) | (r = 1; m ≥ 5) |
| Russia, n=116 | 10,563,038 | 1.560 (3 ÷ 61) | p = 0.21 no DK | **(r = 2; m ≥ 3)** |
| USA, n=283 | 8,391,881 | 1.406 (3 ÷ 283) | p = 0.40 no DK | NaN |
| Germany A, n=187 | 3,431,675 | 1.321 (2 ÷ 55) | p = 0.43 no DK | NaN |
| Germany B, n=187 | 3,431,675 | 4.535 (5 ÷ 15) | **p = 0.0023; 0.0010; 0.0006; 0.0029; 4 DK** | **(r = 4; m ≥ 7)** |
| Argentina, n=236 | 3,058,300 | 1.853 (2 ÷ 33) | p = 0.21 | NaN |
| China, n=153 | 14,230,992 | 1.864 (2 ÷ 22) | p = 0.18 | NaN |
| France, n=39 | 2,193,030 | 1.843 (3 ÷ 39) | p = 0.12 (Paris is close to DK) | Nan |
| Netherlands, n=259 | 767,457 | 1.781 (2 ÷ 48) | p = 0.43 | NaN |
| Japan, n=160 | 8,802,067 | 1.478 (2 ÷ 98) | p = 0.27 | NaN |
| Brazil, n=250 | 11,125,243 | 1.304 (3 ÷ 98) | p = 0.36 | NaN |
| South Africa, n=124 | 2,415,408 | 0.915 (1 ÷ 76) | p = 0.27 | NaN |
| Australia, n=112 | 3,641,422 | 0.902 (1 ÷ 76) | p = 0.18 | NaN |
| Canada, n=183 | 4,753,120 | 0.783 (2 ÷ 86) | p = 0.26 | NaN |

## 4. Testing for Dragon-Kings in agglomeration sizes

We apply the DK-test and U-test to the distribution of agglomeration sizes of two countries (France and USA). These tests have also been performed on many more countries, whose data is found on the website www.citypopulation.de. An urban agglomeration is distinct from a city and is usually defined as an extended city or town area comprising the built-up area of a central place (usually a municipality) and any suburbs linked by continuous urban area. In France, INSEE (the French Statistical Institute) translate it as «Unité urbaine», which means continuous urbanized area. An agglomeration is such that the connected region of dense, predominately urban population is economically and culturally linked to the central city.

### *4.1 France : evidence of one DK (Paris)*

Fig.13 shows the complementary cumulative distribution function CCDF *1-F(x)* in double log-scale for French agglomeration sizes, in 2006, with a total number *n = 217* cities. One can observe that the largest observation (Paris agglomeration) deviates rather significantly from the others and can be suspected as a DK. The choice of the lower threshold *h* above which the power tail should be estimated is not obvious, since the tail shown in Fig.13 has a negative curvature. We chose the lower threshold $h = 173,000$ (leaving 37 observations above this threshold). The ML-estimate of the exponent $b$ corresponding to the power law fit shown in the inset of Fig.13 is found equal to *1.332*.

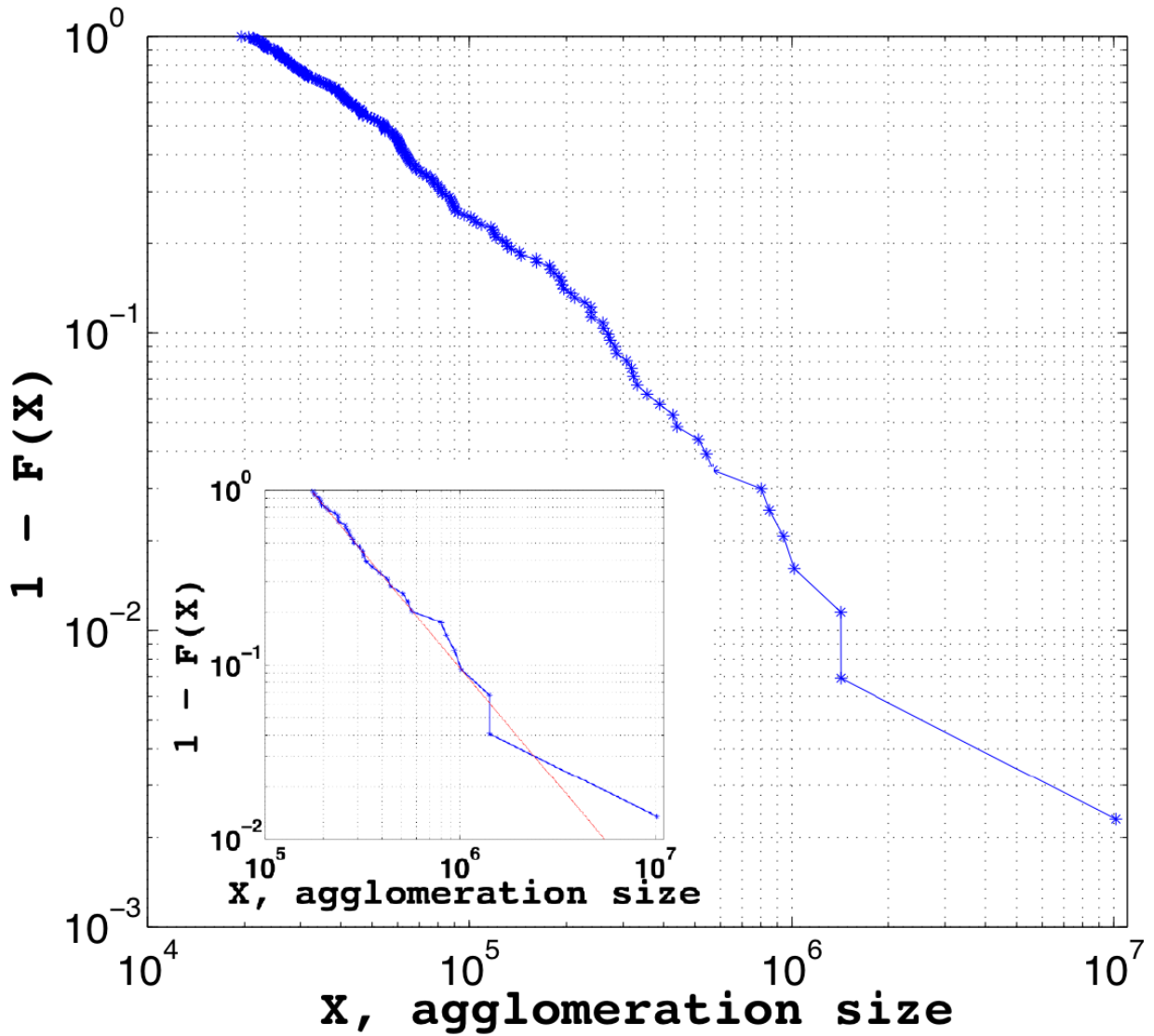

**Figure 13**: Complementary cumulative distribution function (CCDF) of the sizes of the *n=217* largest agglomerations in France (2006). Inset: tail of the CCDF for the *n=37* largest agglomerations with more than *h=173000* inhabitants and fit with a power law (straight line on the graph) whose exponent is found equal to *b=1.332*.

To apply the U-test, we inserted the value $b = 1.332$ into $F(x) = 1 - exp(-b \cdot (x-h))$ and calculated the p-values by equation (10). The result in Fig.14 shows that there are no *p*-values less than 0.10, although the first *p*-value is close to 0.10 (0.144). One can conclude that U-test did not discover any DK, although there is some doubt concerning the maximum observation (Paris agglomeration).

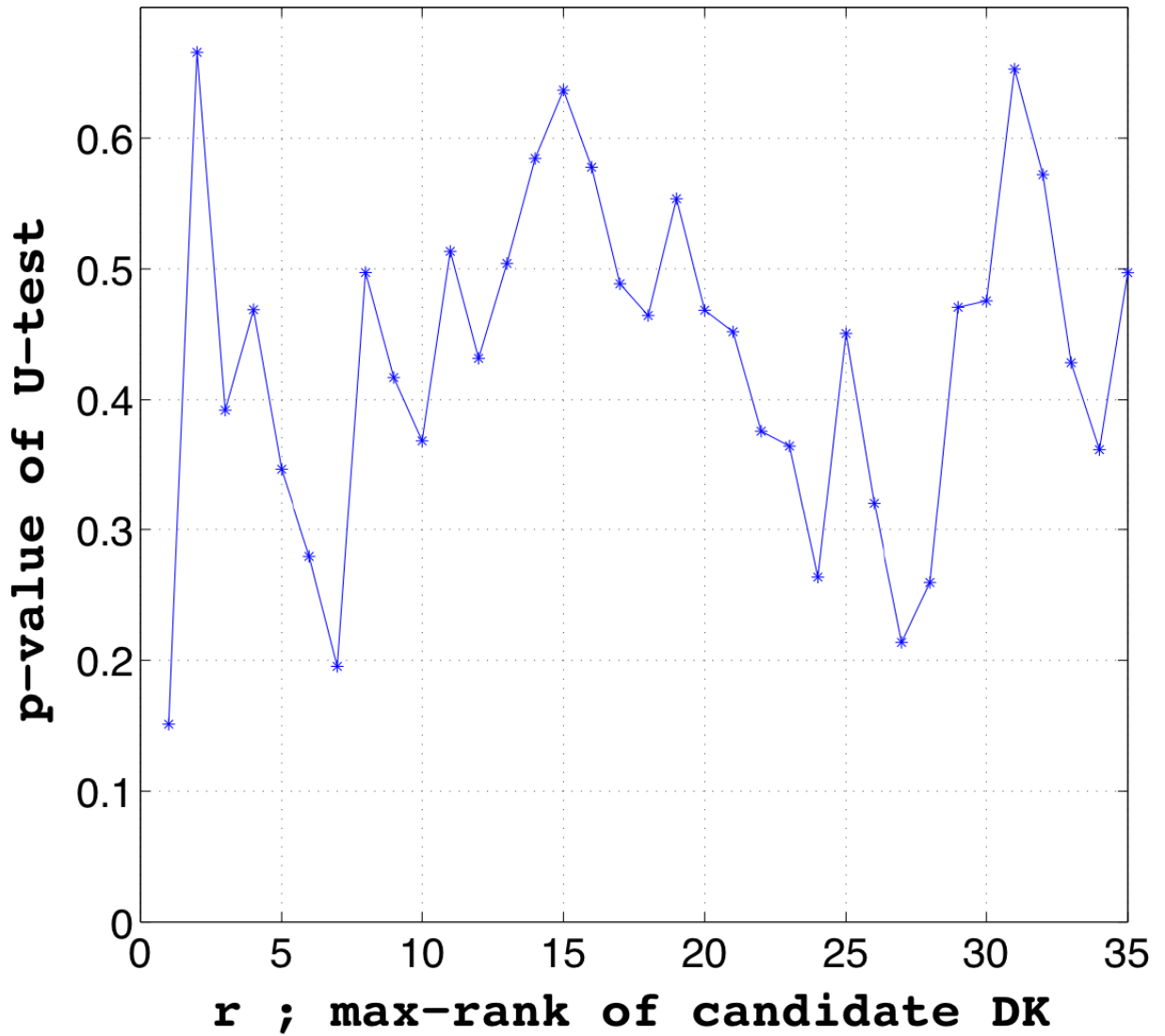

**Figure 14**: *p*-value of the U-test as a function of the maximum rank *r* for candidates DK among the 37 largest observations above the lower threshold *h=173000* for French agglomerations (2006).

To apply the DK test, we took the largest spacing *(x$_{1,37}$ – x$_{2,37}$)* and compared it with the average value of the following ones. Equation (8) with *r=1* (one DK candidate) and a variable number *n$_1$* = *2, 3, ...,30* of the first following ranks yields the *p*-value as a function of *n$_1$* shown in Fig.15. The first observation (Paris) is clearly qualified as a DK by the DK-test, as the corresponding p-value of the largest spacing is equal to 0.007.

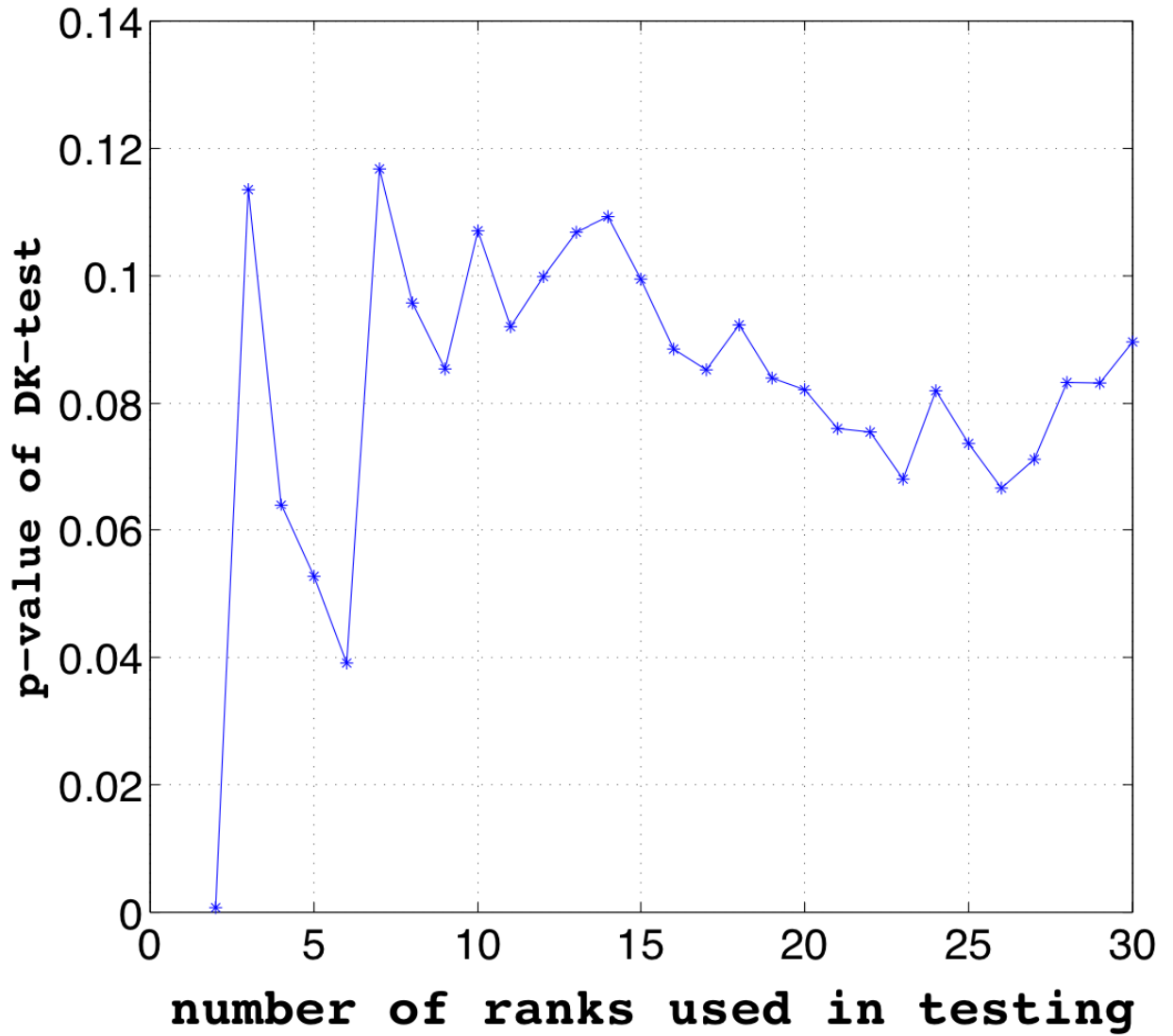

**Figure 15**: *p*-value of the DK-test as a function of the number $n_1$ of ranks used in the testing, for r=1 (Paris as DK candidate for the distribution of French agglomerations, 2006).

*4.2 USA*

Fig.16 shows the complementary cumulative distribution function CCDF *1-F(x)* in double log-scale for US agglomeration sizes, in 2009, with a total number *n = 366* agglomerations. None of the largest observations deviate visually above a linear extrapolation of the CCDF of smaller sizes. The choice of lower threshold *h* is not obvious, since the tail in Fig.16 has a negative curvature. We took first a lower threshold equal to *h = 146000* (leaving 39 observations), a second variant being tried later. The ML-estimate of the exponent *b* corresponding to the power law fit shown in the inset of Fig.16 is found equal to *1.298*.

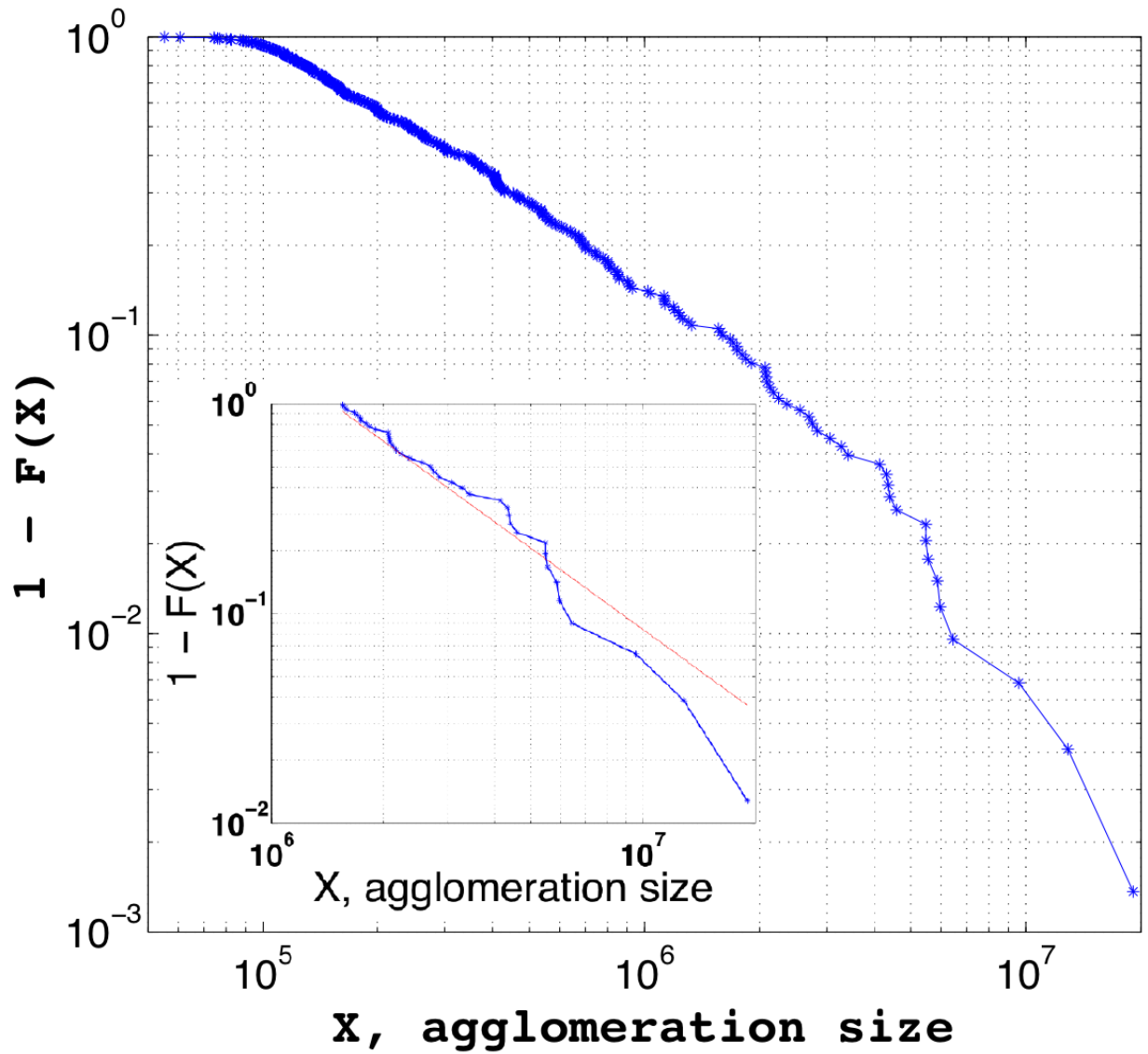

**Figure 16**: Complementary cumulative distribution function (CCDF) of the sizes of the *n=366* largest agglomerations in the USA (2009). Inset: tail of the CCDF for the *n=39* largest agglomerations with more than *h=146000* inhabitants and fit with a power law (straight line on the graph) whose exponent is found equal to *b=1.288*.

To apply the U-test, we inserted the value *b = 1.298* into *F(x) = 1 – exp(-b·(x-h))* and calculated the p-values by equation (10). The result in Fig.17 shows that there are no *p*-values smaller than 0.10, although the 14-th *p*-value is close to 0.10 (0.15). One can conclude that the U-test did not discover any DK.

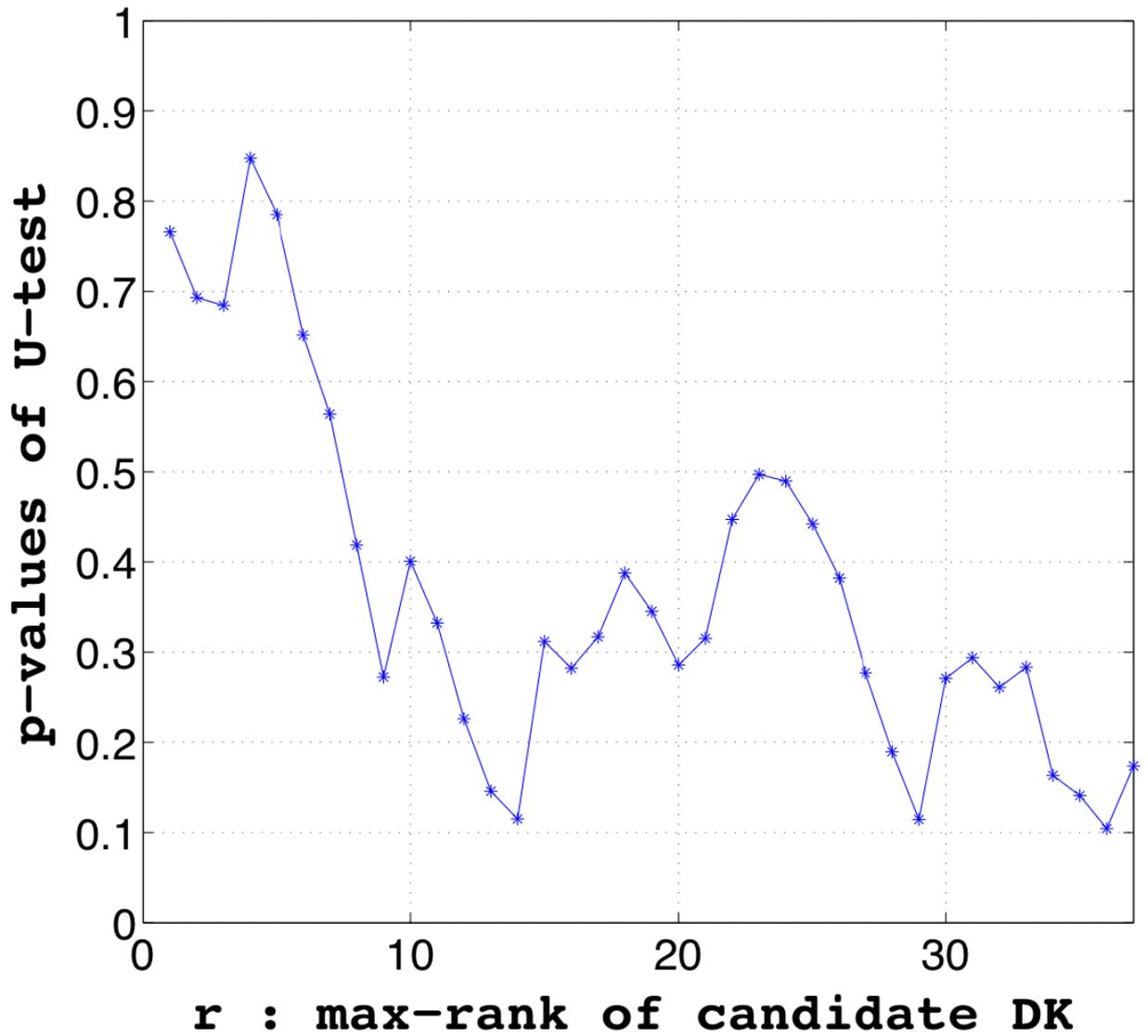

**Figure 17**: *p*-value of the U-test as a function of the maximum rank *r* for candidates DK among the 39 largest observations above the lower threshold *h=146000* for US agglomerations (2009).

To apply the DK test, we took the largest spacing $(x_{1,39} - x_{2,39})$ and compared it with the average value of the following ones. Equation (8) with *r=1* (one DK candidate) and a variable number $n_1$ = *2, 3, …,30* of the first following ranks yields the *p*-value as a function of $n_1$ shown in Fig.18. As all p-values are much larger than 0.1, there is no DK, according to the DK-test.

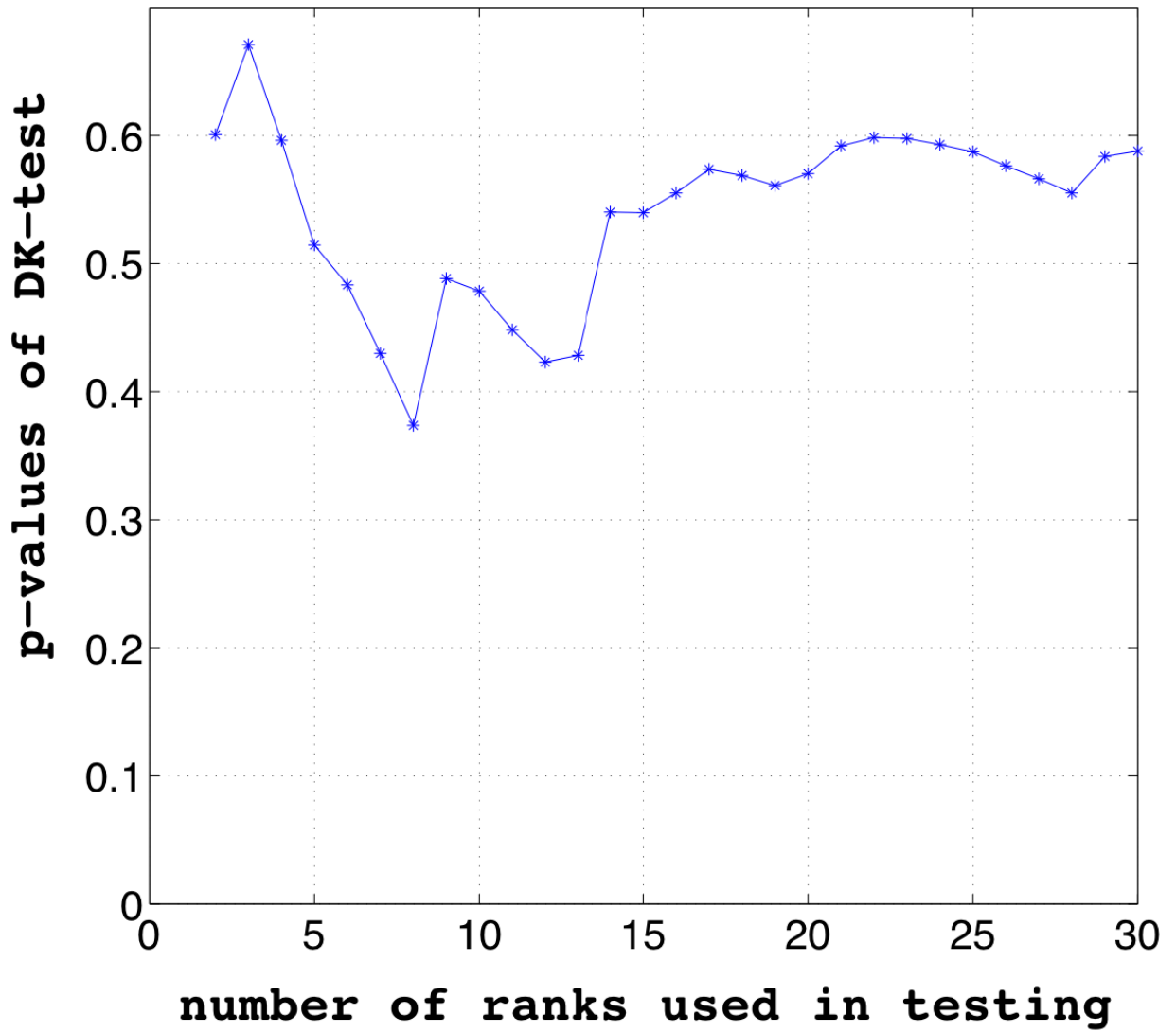

**Figure 18**: *p*-value of the DK-test as a function of the number $n_1$ of ranks used in the testing, for *r=1*, *h=1460000* (New York as DK candidate for the distribution of US agglomerations, 2009).

Let us try identifying a DK by using a second choice for the rank interval in the U and DK tests. A visual inspection of Fig.16 suggests that the 14 largest ranks deviate from the rest of the population. We thus choose *n = 14* corresponding to the threshold *h = 4050000*. We also surmise that there may be r=3 DK among this population of 14 agglomerations. Fitting a power law to the 11 observations in this sample (excluding the 3 potential DK) gives a MLE of the exponent equal to *b = 2.816*, as seen in Fig.19.

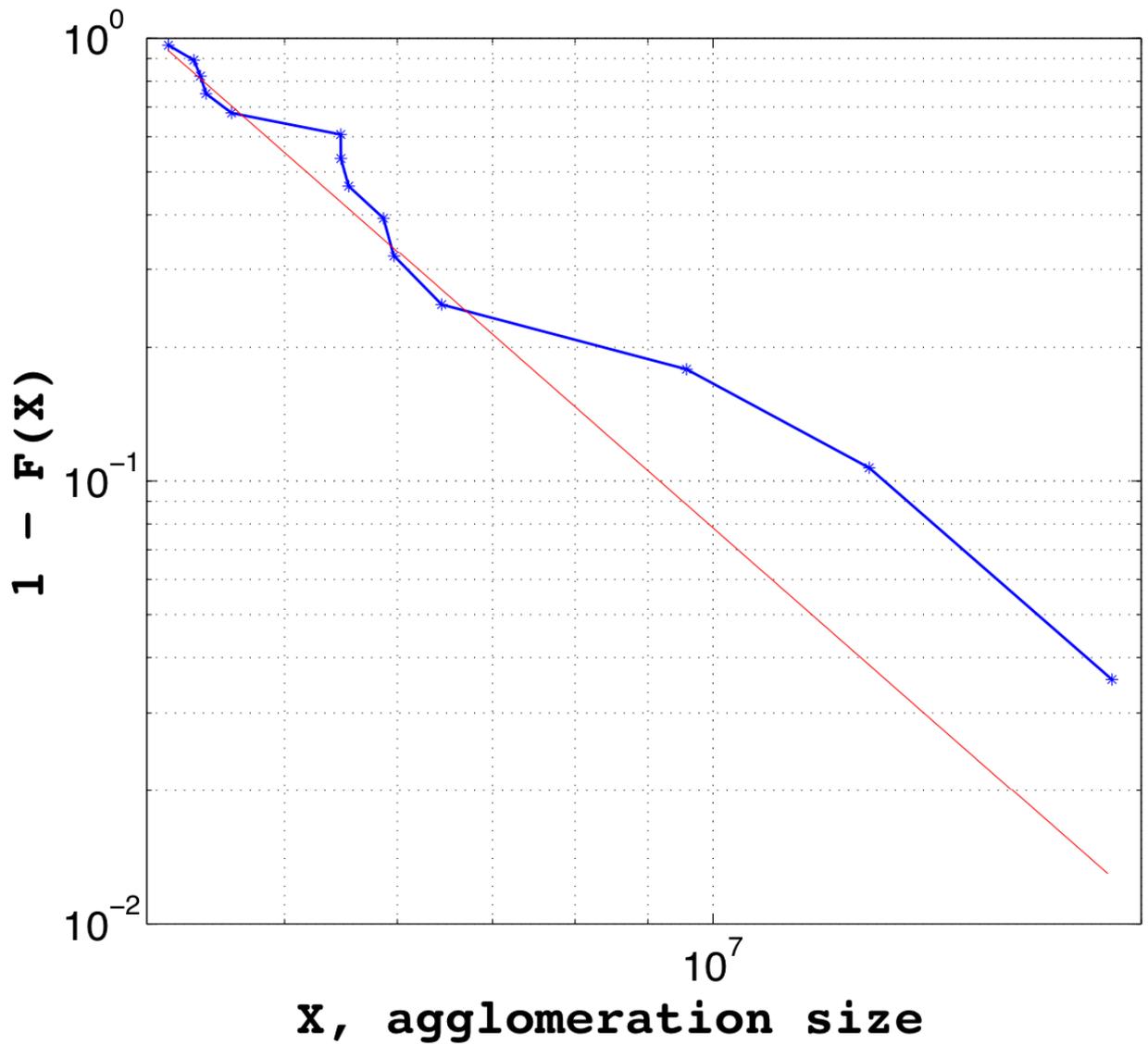

**Figure 19**: Tail of the CCDF for the *n=14* largest agglomerations with more than *h=4050000* inhabitants and fit with a power law (straight line on the graph) on this data set, excluding the 3 largest observations (New York, Los Angeles and Chicago). The corresponding exponent is found equal to *b=2.816*.

To apply the U-test, we inserted the value *b = 2.816* into *F(x) = 1 – exp(-b·(x-h))* and calculated the p-values by equation (10). The result in Fig.20 shows that the qualification of the three largest agglomerations as DK among the 14 largest ones is marginally significant.

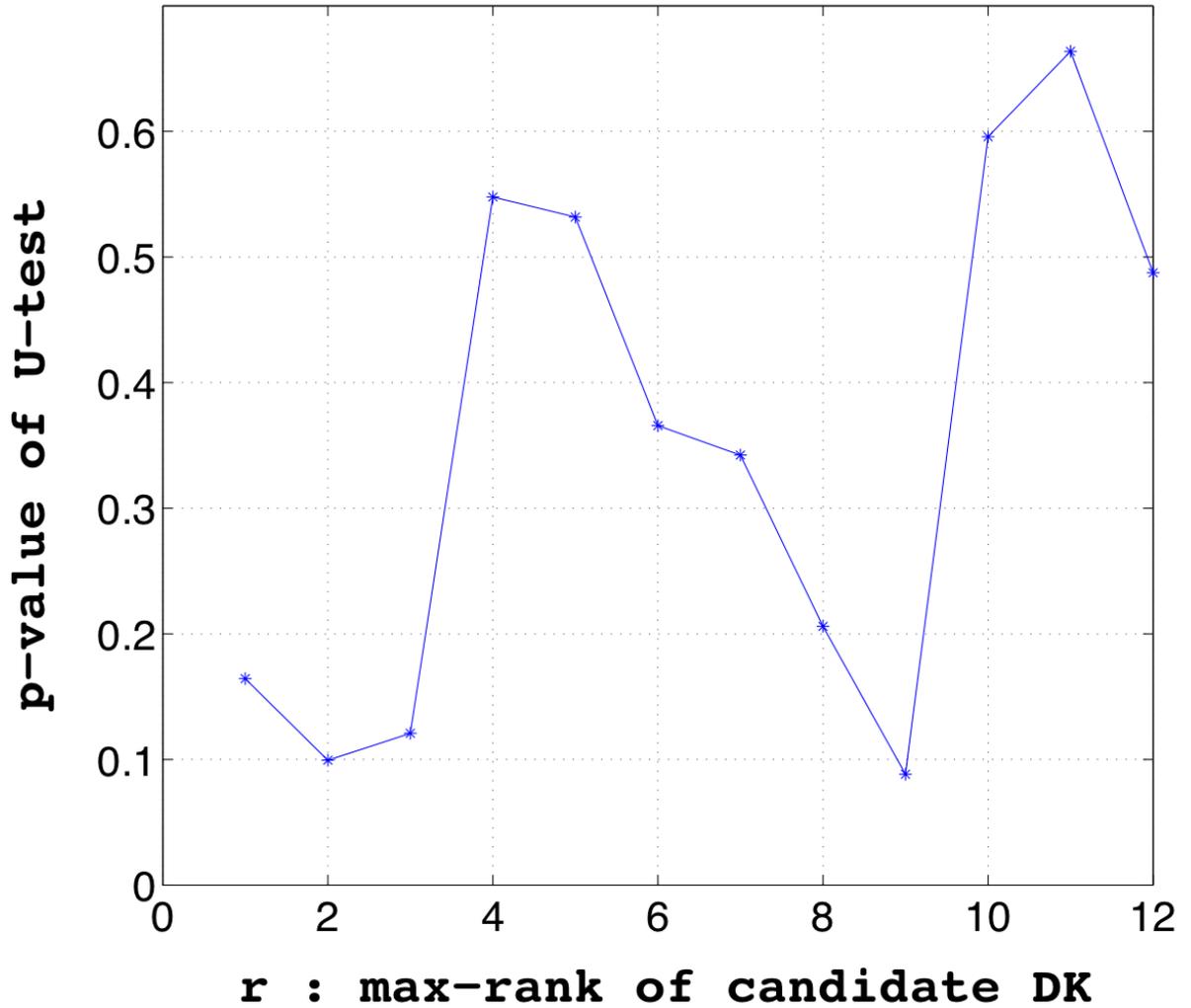

**Figure 20**: *p*-value of the U-test as a function of the maximum rank *r* for candidates DK among the 14 largest observations above the lower threshold *h=4050000* for US agglomerations (2009).

To apply the DK test, we took the three largest spacing $(x_{1,14} - x_{2,14})$, $(x_{2,14} - x_{3,14})$, $(x_{3,14} - x_{4,14})$ and compared them with the average value of the following ones. Equation (8) with *r=3* (three DK candidates) and a variable number $n_1 = 4, 3, ...,12$ of the following ranks yields the *p*-value as a function of $n_1$ shown in Fig.21, which supports that the three largest agglomerations (New York, Los Angeles and Chicago) are DK when compared with the 11 immediately smaller ones.

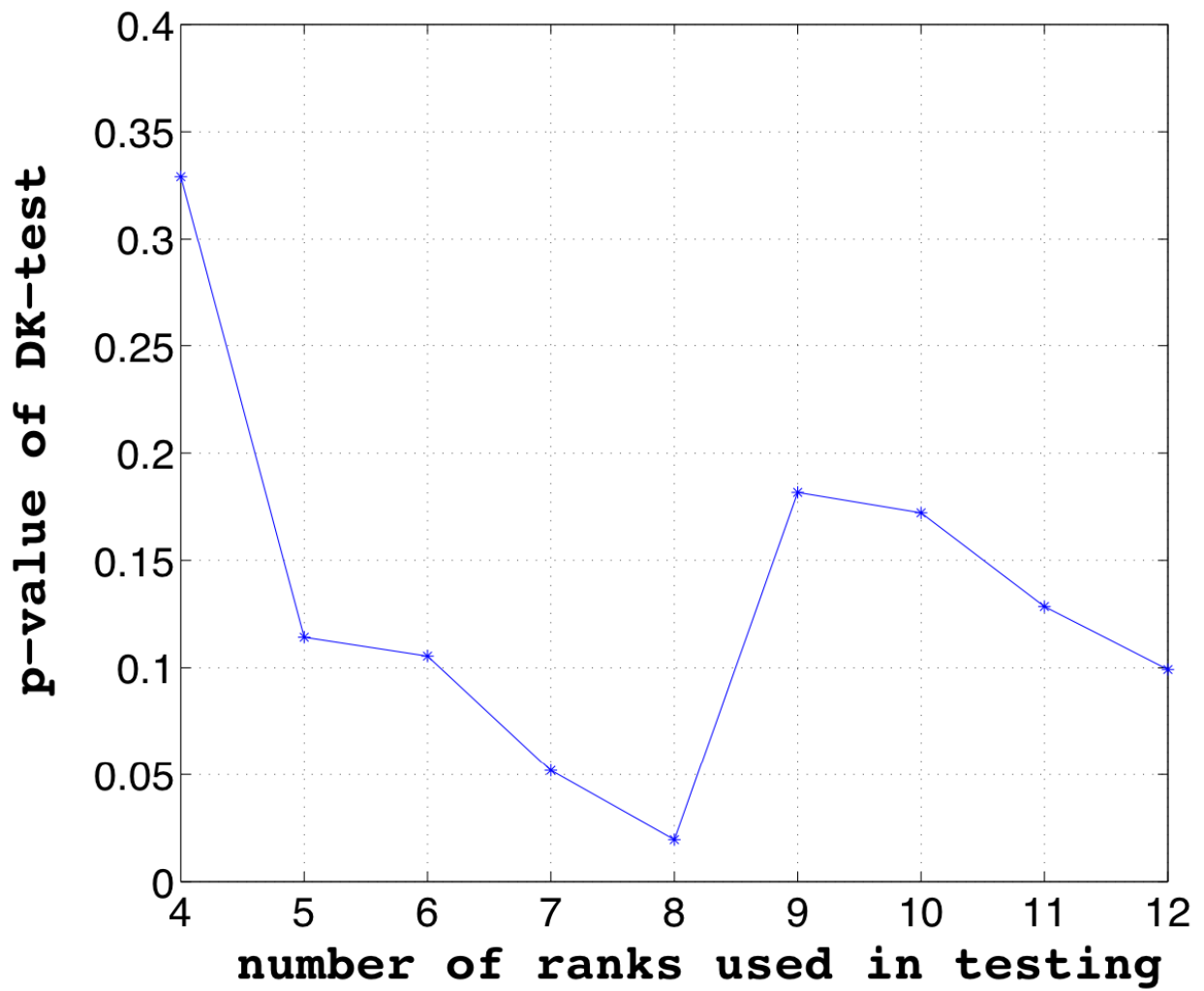

**Figure 21**: *p*-value of the DK-test as a function of the number $n_1$ of ranks used in the testing, for *r=3*, *h=4050000* (New York, Los Angeles and Chicago as DK candidates for the distribution of US agglomerations, 2009).

These results support the notion that the detection of DK depends on what part of the preceding range is taken for comparison. It may therefore be worthwhile to define a DK as ***relative to some m preceding observations***.

*4.3 Summary of results for 6 countries*

Table 3 provide the main summary statistics and synopsis of the results obtained with the U-test and DK-test applied to the distribution of agglomerations in 6 different countries.

Table 3: Agglomeration size statistics. M = number of spacing in denominator of (7) giving p-value smaller than 0.10;   r = number of spacing in numerator of (7) ; n = total number of agglomerations considered in the tests.

| country | max size | b, Pareto exponent | min p-value of U-test | DK-test (r; m) giving |
|---|---|---|---|---|

| | | (ranks used for estimation) | among first 10 ranks | p<0.10 |
|---|---|---|---|---|
| USA A, n=366 | 19,069,796 | 1.288 (2 ÷ 39) | p = 0.27 | NaN |
| USA B, n=366 | 19,069,796 | 2.816 (4 ÷ 14) | **p = 0.16; NY p=0.099;LosAngeles p = 0.12; Chicago** | **(3; 6≤m≤8)** |
| Great Britain, n=260 | 8,278,251 | 0.981 (2 ÷ 116) | p = 0.52 | NaN |
| France, n=217 | 10,142,977 | 1.332 (2 ÷ 37) | **p = 0.145 (Paris)** | **(r = 1; m ≥4)** |
| Germany, n=195 | 4,694,686 | 0.995 (1 ÷ 56) | p = 0.34 | NaN |
| Canada, n=143 | 5,113,149 | 0.744 (4 ÷ 88) | p = 0.41 | **(r=3; m=5)** |
| Brazil, n=35 | 19,672,582 | 1.339 (4 ÷ 15) | p = 0.20 | **(2; 6≤m≤8)** |

## 5. Conclusion

We have presented two new statistical tests, the U-test and the DK-test, that are aimed at identifying the existence of even a single anomalous event in the tail of the distribution of just a few tens of observations. The U-test is based a specific measure of deviations from the null hypothesis, which can be either an exponential or power law distribution. It uses the maximum likelihood estimation of the exponent characterizing the null hypothesis and its p-value is obtained from a simple formula in terms of the normalized incomplete beta functions. The DK-test also uses a null hypothesis, which can be either an exponential or power law distribution. It amounts to comparing the rescaled spacing between the highest ranks and the lowest ranks. The statistic is chosen so that its p-value is independent of the exponent characterizing the null hypothesis.

We have illustrated the application of these two tests on the distributions of cities and of agglomerations in a number of countries. We have presented examples where Dragon-Kings (DK) are clearly identified, as well as cases in which we fail to detect any anomaly and the null hypothesis of a power law distribution holds.

We expect that the U-test and DK-test will become useful standard additions to the toolbox of scientists interested in characterizing the tails of distributions that are found in all fields of natural and social sciences. Our purpose has been to demonstrate that the characterization of power laws is not sufficient. We hope to have shown that there may be anomalies or deviations in the extreme tail that can be quantified.